\documentstyle[preprint,aps,floats,epsfig]{revtex}

\tightenlines

\def\appendix{\par
 \setcounter{section}{0}
 \setcounter{subsection}{0}
 \def\thesection{Appendix \Alph{section}}
 \def\thesubsection{\Alph{section}.\arabic{subsection}}
 \def\theequation{\Alph{section}.\arabic{equation}}
 \setcounter{equation}{0}}
\newcommand{\be}{\begin{equation}}
\newcommand{\ee}{\end{equation}}
\newcommand{\bear}{\begin{eqnarray}}
\newcommand{\eear}{\end{eqnarray}}

\begin{document}
\draft
\title{Resummation of the hadronic tau decay width with
the modified Borel transform method}

\author{Gorazd Cveti\v c$^1$\footnote{cvetic@fis.utfsm.cl},
Claudio Dib$^1$\footnote{cdib@fis.utfsm.cl}, 
Taekoon Lee$^2$\footnote{tlee@muon.kaist.ac.kr},
and Iv\'an Schmidt$^1$\footnote{ischmidt@fis.utfsm.cl}}
\address{$^1$Department of Physics, 
Universidad T\'ecnica Federico Santa Mar\'{\i}a,     
Valpara\'{\i}so, Chile\\
$^2$Department of Physics, KAIST, Daejon 305-701, Korea}
\date{\today}

\maketitle

\begin{abstract}
A modified Borel transform of the Adler function is used
to resum the hadronic tau decay width ratio. 
In contrast with the ordinary Borel transform, the integrand
of the Borel integral is renormalization--scale invariant.
We use an ansatz which explicitly accounts for the structure of 
the leading infrared renormalon. Further, we use judiciously
chosen conformal transformations for the Borel variable, in
order to map sufficiently away from the origin the other ultraviolet and
infrared renormalon singularities. In addition, we apply Pad\'e
approximants for the corresponding truncated perturbation
series of the modified Borel transform, in order to further
accelerate the convergence. Comparing the results with the
presently available experimental data on the tau hadronic
decay width ratio, we obtain $\alpha_{\text{s}}(M^2_{\text{z}})
=0.1192 \pm 0.0007_{\rm exp.} \pm 0.0010_{\rm EW+CKM} \pm 
0.0009_{\rm th.} \pm 0.0003_{\rm evol.}$.
These predictions virtually agree with those of our previous
resummations where we used ordinary 
Borel transforms instead.\\
\noindent
PACS number(s): 13.35.Dx, 11.15.-q, 11.15.Tk

\end{abstract}
\pacs{}
\section{Introduction}
\label{sec:introduction}

Extensive perturbative calculations in QCD have made available
the truncated perturbation series (TPS) of various observables
to the next--to--next--to--leading order (NNLO), i.e., including
the terms $\sim\!\alpha_s^3$. There is a longstanding problem
of how to extract (resum), in a reasonable manner, the values of
such observables $S(\alpha_s)$ as functions of the strong
coupling parameter $\alpha_s^{\overline {\rm MS}}(M^2_{\rm z})$.
Many of the resummation methods are based solely on the
available TPS $S_{[3]}(\alpha_s)$. Some of these methods
eliminate the (unphysical) renormalization--scale (RScl)
and renormalization--scheme (RSch) dependence of the TPS
by fixing the RScl and RSch in the TPS itself. Among them
are: the RScl fixing of Brodsky, Lepage and Mackenzie (BLM),
motivated by large--$n_f$ arguments \cite{Brodsky:1983gc}; 
the principle of minimal
sensitivity (PMS) \cite{PMS}; and effective charge method (ECH)
\cite{ECH,Kataev:1982gr,Gupta}. Some of the more recent methods are:
the approaches using ``commensurate scale relations'' \cite{csr};
ECH--related approaches 
\cite{ECHrel,Maxwell:1998yw,Maxwell:2001uv,Korner:2001xk};
a method using an analytic form of the coupling parameter 
\cite{AQCD};
a method using expansions in the two--loop coupling parameter
\cite{Kourashev};
a method which disentangles the running coupling and
conformal effects (skeleton coefficients) \cite{Brodsky:2001cr};
methods using conformal transformations either 
for the coupling parameter \cite{Solovtsov}
or for the Borel expansion parameter \cite{Lee:1997yk,Caprini:1999wg};
the method of Pad\'e approximants (PA's) \cite{Pade,Gardi,Steele:1998ma};
an RScl--invariant \cite{Cvetic1} and RScl-- and RSch--invariant
extension of the PA-approach \cite{Cvetic:2001mz}.

Some of the aforementioned approaches allow for 
incorporation of additional, nonperturbative, information
on the observable, e.g., the information on the location of 
the dominant infrared (IR) renormalon singularity \cite{Cvetic:2001mz}
via the fixing of the remaining free parameter in the
approximant, or the information on the location of
various renormalon singularities via leading--$\beta_0$ resummation
\cite{Maxwell:1998yw,Maxwell:2001uv}.
Full information on the known leading IR renormalon structure
can be incorporated in a natural way 
through an explicit ansatz, in the approaches using
Borel transformations of the considered or of the
associated observable (see, e.g., \cite{Cvetic:2001sn}).
Incorporation of the available leading renormalon
information appears to influence significantly
the numerical predictions, particularly for
low--energy (low--$Q^2$) QCD observables, 
through a more accurate description of the Borel
amplitude in the region between the origin and 
the leading IR renormalon singularity.
This method has an advantage over the widely used (unresummed)
Operator Product Expansion (OPE) approach
\cite{OPE}, in which the perturbative contributions are
usually taken as TPS, and additional nonperturbative 
terms $\propto\!1/Q^{2 n}$ appear (cf.~\cite{Braaten:1992qm} and
references therein).
Since the perturbation series is divergent, 
the latter approach has the problem that
there is no natural separation between perturbative contributions
and power--suppressed nonperturbative terms, 
a problem which does not occur in the former method. 
However, the Borel--resummed value for an observable, with
the renormalon information incorporated in it 
(e.g., via an explicit ansatz),
in principle still does not represent the full value
of the observable, because power--suppressed terms should
be added to it.
Dispersive models
(\cite{dispersive} and references therein) 
contain and predict the OPE--type power--suppressed terms.
Recently, it has been argued 
\cite{Lee:2001ws} that the power--suppressed terms can be 
obtained from the knowledge of the perturbation series, or parts thereof, 
and of the IR renormalon structure of the 
observable.

In all these resummation approaches, for obvious
reasons, it is highly preferable
to consider QCD observables whose values are
presently known at a high precision level. 
Presently, high precision experimental data on $\tau$ lepton
decay widths 
\cite{Barate:1998uf,Barate:1999hj,Davier:1999xy,Abbiendi:2000cq,Richichi:1999bc,Davier2}
are available. The non--strange hadronic $\tau$ decay width ratio 
$R_{\tau}(\triangle S\!=\!0)$
can be obtained by using constrained fit values
\cite{Groom:2000in} of the (basis modes) leptonic branching ratios 
$B_e \equiv B( \tau^- \to e^- {\overline \nu}_e \nu_{\tau})
= (17.83 \pm 0.06) \times 10^{-2}$ and 
$B_{\mu} \equiv B( \tau^- \to \mu^- {\overline \nu}_{\mu} \nu_{\tau})
= (17.37 \pm 0.07) \times 10^{-2}$ of $\tau$,
and subtracting the strangeness--changing part \cite{Davier2}
$R_{\tau}(\triangle S\!\not=\!0) = 0.1630 \pm 0.0057$  
\begin{eqnarray}
R_{\tau}(\triangle S\!=\!0)  &\equiv& 
\frac{ \Gamma (\tau^- \to \nu_{\tau} {\rm hadrons} (\gamma) )}
{ \Gamma (\tau^- \to \nu_{\tau} e^- {\overline {\nu}_e} (\gamma))}
- R_{\tau}(\triangle S\!\not=\!0)
\label{Rtaudef}
\\
& = & \frac{(1 - B_e - B_{\mu} )}{B_e} - R_{\tau}(\triangle S\!\not=\!0)
= 3.4713  \pm 0.0171 \ . 
\label{Rtauexp}
\end{eqnarray}
The above ratio is a QCD observable at relatively low
momenta $\sqrt{q^2} \sim m_{\tau} \approx 1.8$ GeV, 
experimentally known to high precision,
thus presenting an experimental challenge to the theory. 
The challenge consists in predicting the strong
coupling constant $\alpha_s^{\overline {\rm MS}}(M_Z^2)$ so that
the theoretical uncertainty $(\delta \alpha_s)_{\rm th.}$, 
which partly originates from the uncertainty of the method of resummation
and partly from the uncertainty of the associated Adler function, 
is smaller or comparable
to the uncertainty $(\delta \alpha_s)_{\rm exp.}$ originating from
the small experimental uncertainty $\delta R_{\tau} \approx \pm 0.017$
given above. 

There is now a wealth of theoretical results 
\cite{Braaten:1992qm,Lam:1977cu,Schilcher:1984ae,Narison:1988ni,rtau3,Pich:1990um,Pivovarov:1992rh,rtau7}
available on
the observable (\ref{Rtauexp}), and, in particular, on the
associated Adler function $D(Q^2)$ -- perturbative as well as
nonperturbative. The main QCD theoretical problem connected with
the observable (\ref{Rtauexp}) is that the pertaining
momenta are somewhat low $|Q|\!\sim\!m_{\tau}$ ($\approx 1.8$ GeV),
and so the relevant coupling parameter $\alpha_s(Q^2)$
in the perturbation expansion is large. It is thus
important to take into account, in any resummation procedure
for $R_{\tau}$ and/or $D(Q^2)$,
not just the known perturbative coefficients, but
also a significant part of the nonperturbative
information, i.e., the leading infrared renormalon.
On the other hand, the nonperturbative contributions
to $R_{\tau}$ that are represented by
power--suppressed OPE terms (apart from the well known
quark mass contributions) have been shown to be consistent
with zero in the ALEPH analysis \cite{Barate:1998uf}.

One version of this program was carried out in our previous
work \cite{Cvetic:2001sn}. There we first used the known information
on the leading infrared (IR) renormalon and on the corresponding
$1/Q^2$-suppressed term of the OPE for the
Adler function $D(Q^2)$, in order to predict the 
${\cal {O}}(\alpha_s^4)$
coefficient $d_3(Q^2)$ of $D(Q^2)$. Using this prediction, and
judiciously chosen conformal transformations, we 
resummed $R_{\tau}$ by employing an ordinary Borel transform 
${\widetilde D}(b)$ of the Adler function with an ansatz
which explicitly accounted for the leading IR renormalon
structure. Comparing the obtained expression with the
experimental values (\ref{Rtauexp}), we obtained definite
predictions for the strong coupling parameter 
$\alpha_s(M^2_{\rm z}) = 
0.1193 \pm 0.0007_{\rm exp.} \pm 0.0010_{\rm EW+CKM} \pm
0.0009_{\rm meth.} \pm 0.0003_{\rm evol.}$.
We fixed the RScl by the principle of minimal
sensitivity (PMS) applied to the approximant, 
and by choosing the ${\overline {\rm MS}}$ RSch.

In the present work instead, in order to obtain
a cross--check of the predictions, we employ 
not ordinary Borel transforms but modified Borel transforms of $D(Q^2)$.
Such Borel transforms were introduced by Grunberg \cite{Grunberg:1993hf},
on the basis of the modified Borel transforms of Ref.~\cite{modBT}.
One attractive feature of Grunberg's transforms is
that the integrand in the Borel integral is
RScl--independent, in contrast to the case of the
ordinary Borel transforms. Another interesting feature is that 
they represent integral transformations of a significantly different
form than the ordinary Borel transformation, and have therefore
a different singularity structure. Therefore, their application
to the hadronic tau decay width ratio and the subsequent
extraction of the prediction for $\alpha_s(M^2_{\rm z})$ could
represent a powerful cross--check of the results of previous
work \cite{Cvetic:2001sn} based on ordinary Borel transformations.

In Sec.~\ref{sec:basic} 
we recall the known basic theoretical formulas for
$R_{\tau}(\triangle S\!=\!0)$ and
the associated Adler functions, as well as
the reduction to the massless QCD observable $r_{\tau}$.
In Sec.~\ref{sec:modBT}, we present the modified Borel transforms
${\overline D}(b)$ of the massless Adler function $D(Q^2)$. 
We refer to the Appendix for details about ${\overline D}(b)$.
In Sec.~\ref{sec:resummation} we perform the resummation for
$r_{\tau}$, by contour integration of $D(Q^2)$ 
in the complex momentum plane,
explicitly accounting for the leading IR renormalon
structure of ${\overline D}(b)$ through an ansatz,
choosing conformal transformations of the
Borel variable $b$ to map away other singularities of ${\overline D}(b)$, 
and employing Pad\'e approximants.
In Sec.~\ref{sec:predictions}
we compare the obtained expressions with the experimental
results and determine $\alpha_s(m^2_{\tau})$ and
$\alpha_s(M^2_{\rm z})$. We further estimate the theoretical
uncertainties of the prediction.
Sec.~\ref{sec:summary} contains a summary 
and brief discussion of the differences between our results
and those of other analyses of $R_{\tau}$.

\section{The known basic formulas, reduction to massless QCD}
\label{sec:basic}

The restriction $\triangle S\!=\!0$ in Eq.~(\ref{Rtauexp}) 
means that only hadrons with
quarks $u$ and $d$ are produced. Thus the observable
(\ref{Rtauexp}) is already close to being massless. This fact
removes some of the complications in the theoretical analysis.

The ratio (\ref{Rtauexp}) can be expressed, via the application of a
variant of the optical theorem, and the subsequent use of
Cauchy's theorem and integration by parts, as a contour integral
in the complex momentum plane
(see, for example, Refs.~\cite{Braaten:1992qm,rtau7}):
\begin{eqnarray}
\lefteqn{
r_{\tau}(\triangle S\!=\!0) \equiv 
\frac{ R_{\tau}(\triangle S\!=\!0) }
{ 3 |V_{ud}|^2 (1 + {\delta}_{{\rm EW}}) } -
(1 + \delta_{\rm EW}^{\prime} )
}
\label{rtgen0}
\\
&& = ( - \pi {\rm i} ) \int_{|s|=m^2_{\tau}}
\frac{ds}{s} \left( 1\!-\!\frac{s}{m^2_{\tau}} \right)^3
\left[ \left( 1 + \frac{s}{m^2_{\tau}} \right) D^{\rm L\!+\!T}(-s)
+ \frac{4}{3} D^{\rm L}(-s) \right] - 1 \ .
\label{rtgen}
\end{eqnarray}
Here we factored out, for convenience, the square of the
Cabibbo--Kobayashi--Maskawa (CKM) matrix element $|V_{ud}|$,
the electroweak (EW) correction parameter
$\delta_{\rm EW} = 0.0194 \pm 0.0050$ 
\cite{Braaten:1990ef,Marciano:1988vm,Braaten:1992qm},
and the residual EW correction parameter
$\delta^{\prime}_{\rm EW} = 0.0010$ \cite{Braaten:1990ef}.
The contour integration in Eq.~(\ref{rtgen})
is counterclockwise in the complex $s$--plane,
and the general Adler functions $D^{\rm L\!+\!T}$ and $D^{\rm L}$
are related to the ${\rm V}\!+\!{\rm A}$
current--current correlation functions as:
\begin{eqnarray}
D^{\rm L\!+\!T}(-s) & = & - s \frac{d}{ds} \sum_{J=0,1}
\left( \Pi_{ud,V}^{(J)}(s) +\Pi_{ud,A}^{(J)}(s) \right) \ ,
\label{DLT} 
\\
D^{\rm L}(-s) & = & \frac{s}{m^2_{\tau}} \frac{d}{ds}
\left[ s \left( \Pi_{ud,V}^{(0)}(s) +\Pi_{ud,A}^{(0)}(s) \right) \right] \ .
\label{DL}
\end{eqnarray}
Here, $J$ is the spin of the hadronic system in its
rest frame (L: $J\!=\!0$; T: $J\!=\!1$); $\Pi_{ud,V/A}^{(J)}$
are the components in the Lorentz decomposition 
\begin{equation}
\Pi^{\mu \nu}_{ud,V/A}(q) = ( - g^{\mu \nu} q^2 + q^\mu q^\nu )
\Pi^{(1)}_{ud,V/A}(q^2) + q^\mu q^\nu \Pi^{(0)}_{ud,V/A}(q^2) 
\label{Ldecomp}
\end{equation}
of the two--point
correlation functions $\Pi^{\mu \nu}_{ud,V/A}$
of the vector 
$V_{ud}^{\mu}\!=\!{\bar d} \gamma^{\mu} u$
and axial--vector
$A_{ud}^{\mu}\!=\!{\bar d} \gamma^{\mu} \gamma_5 u$
(color--singlet) currents
\begin{eqnarray}
- {\rm i} \Pi^{\mu \nu}_{ud,V}(q) &=& 
\int d^4 x \; e^{{\rm i} q\; \cdot\; x} 
\langle 0 | {\rm T} \lbrace V_{ud}^{\mu}(x) V_{ud}^{\nu}(0)^{\dagger} 
\rbrace | 0 \rangle \ ,
\label{Vcorr}
\\
- {\rm i} \Pi^{\mu \nu}_{ud,A}(q) &=& 
\int d^4 x \; e^{{\rm i} q \; \cdot \; x} 
\langle 0 | {\rm T} \lbrace A_{ud}^{\mu}(x) A_{ud}^{\nu}(0)^{\dagger} 
\rbrace | 0 \rangle \ ,
\label{Acorr}
\end{eqnarray}
In the massless quark limit ($m_{u,d} \to 0$), $D^{\rm L}(s)$ vanishes
and the vector and axial--vector contributions to
$D^{\rm L\!+\!T}$ become equal in perturbation and
$D^{\rm L\!+\!T}(-s) \to (1 + D(-s))/(2 \pi^2)$,
where $D(Q^2)$ is the canonically normalized massless
Adler function with the perturbative expansion\footnote{
The ($ud$) Adler functions $D^{\rm L+T}$ and $D^{\rm L}$
(\ref{DLT})--(\ref{DL}) usually include, by convention, the 
additional CKM factor $|V_{ud}|^2$ (e.g., see Refs.~\cite{rtau7}).}
\begin{equation}
D(Q^2)= a \left[ 1 + \sum_{n=1}^{\infty} d_n a^n \right] \ .
\label{Dexpan}
\end{equation}
Here, $a\!=\!{\alpha}_s(\mu^2; c_2, c_3, \ldots)/\pi$ is the QCD
couplant at the renormalization scale (RScl) $\mu^2$ and 
in the renormalization scheme (RSch) characterized by the
coefficients $c_j$ ($j \geq 2$) in the beta function
\begin{equation}
\mu^2\frac{d}{d \mu^2} a \equiv \beta(a)
= - \beta_0 a^2 [1 + c_1 a + c_2  a^2 +\cdots] \ .
\label{aRGE}
\end{equation}
Here, $\beta_0\!=\!(11\!-\!2 n_f/3)/4$ and 
$c_1\!=\!(102\!-\!38 n_f/3)/(16 \beta_0)$ 
are two universal constants which depend
only on the number of active quark flavors $n_f$.
The Adler functions are quasiobservables, in the sense that
they are independent of the RScl and RSch.

In order to apply the massless QCD analysis to
$r_{\tau}(\triangle S\!=\!0)$ (\ref{rtgen}), we have to
subtract from it the quark mass ($m_{u,d} \not= 0$)
contributions. This can be carried out \cite{Braaten:1992qm}
within an operator product expansion (OPE).
The largest quark mass contributions are 
quark condensate terms of dimension $d\!=\!4$ 
($\propto\!1/m^4_{\tau}$)
\begin{eqnarray}
\lefteqn{
\delta r_{\tau}(\triangle S\!=\!0)_{m_{u,d}\not=0}
\approx 16 \pi^2 
\frac{ (m_u\!+\!m_d)\langle {\bar q} q \rangle }{m^4_{\tau}}
\left[ 1 + \frac{23}{8} 
\left( \frac{ \alpha_s(m^2_{\tau}) }{\pi} \right)^2 \right]
}
\label{drtmud1}
\\
&\approx & - \frac{16 \pi^2 f^2_{\pi} m^2_{\pi}}{m^4_{\tau}}
\left[ 1 + \frac{23}{8} 
\left( \frac{ \alpha_s(m^2_{\tau}) }{\pi} \right)^2 \right]
\approx - 0.00265 \times (1 + 0.03) \approx - 0.0027 \ .
\label{drtmud2}
\end{eqnarray}
In Eq.~(\ref{drtmud1}) we denoted 
$\langle {\bar q} q \rangle \equiv \langle {\bar u} u \rangle
\approx \langle {\bar d} d \rangle$. Here renormalization scale
can be taken to be
$\mu \approx m_{\tau}$. In Eq.~(\ref{drtmud2}) we used
the PCAC relation $(m_u\!+\!m_d)\langle {\bar q} q \rangle
\approx - f^2_{\pi} m^2_{\pi}$ ($f_{\pi}\!=\!92.4\pm0.3$ MeV;
$m_{\pi^-}\!=\!139.6$ MeV). There are corrections to this
relation and to expression (\ref{drtmud2}) of the order 
$\sim\!m^2_{u,d}/m^2_{\tau}$, i.e., of the order of the OPE
$d\!=\!2$ terms which can reach, at most, values $\sim\!10^{-4}$.
In obtaining the numerical value in Eq.~(\ref{drtmud2}), 
we further used $m_{\tau}\!=\!1777$ MeV and
$\alpha_s(m^2_{\tau},{\overline {\rm MS}}) \approx 0.32$.

The OPE approach of \cite{Braaten:1992qm} includes other 
power--suppressed nonperturbative terms that contribute to $r_{\tau}$ 
but do not stem from quark masses:
the $d=4$ gluon condensate term, and the $d=6$ term.
The latter term could be large, but it has also comparably
large uncertainties \cite{Braaten:1992qm}. The gluon condensate
contribution to $r_{\tau}$ 
in the OPE is $\alpha_s^2$--suppressed.
The ALEPH analysis \cite{Barate:1998uf}
indicates that these $d=4,6$ nonperturbative contributions
to $r_{\tau}$ are consistent with the value zero,
$\delta r_{\tau}({\rm NP}; m_{u,d}\!=\!0) = 0.000 \pm 0.004$.
We should keep in mind, however, that the ALEPH analysis
assumed that the part of the associated Adler function 
which has no power--suppressed terms is
a (${\rm N}^3 {\rm LO}$) TPS, while we will perform
resummations of this part by taking into account 
its renormalon singularity structure. Nonetheless,
we consider the ALEPH analysis as indicative that
even in our resummation framework the 
power--suppressed OPE--type terms in $r_{\tau}$ (apart from
the quark mass terms) are either consistent with
zero or very small.
Therefore, we will ignore in our analysis of $r_{\tau}$
any OPE power--suppressed nonperturbative terms other than those in
Eqs.~(\ref{drtmud1})--(\ref{drtmud2}).

We thus regard, in our framework,
as nonperturbative massless contributions only those 
contributions which appear as a consequence of explicit 
IR renormalon structure of the Borel transforms in the resummation.
For example, the leading IR renormalon of the Adler function $D(Q^2)$
(which we will account for) gives contributions to $D(Q^2)$
which can be partially represented as a $d\!=\!4$ power--suppressed 
term $\propto\!1/Q^4$. This, however, does not necessarily mean that
there is no additional, genuine OPE--type $d\!=\!4$ term 
($\propto\!\langle a G G \rangle/Q^4$) in $D(Q^2)$.
The uncertainties of the OPE $d\!=\!4,6$ massless terms as given 
by ALEPH are large enough to accommodate the possibility of
significant nonzero values of these terms. For example, 
$\langle a G G \rangle = 0.001 \pm 0.015 \ {\rm GeV}^4$.
Toward the end of Sec.~\ref{sec:predictions} we will
briefly discuss how the latter uncertainties would influence
the final prediction for $\alpha_s$ in our analysis.

The ALEPH analysis further assumed that massless $d\!=\!2$ terms 
($\propto\!1/Q^2$) are not present in the Adler function
(and thus in $r_{\tau}$), as suggested by the OPE.
Such terms were suggested by the authors of
Refs.~\cite{Chetyrkin:1999yr,Huber:1999ug} 
as an effective tachyonic gluon mass 
contribution reflecting nonperturbative short--distance QCD.
However, the authors of Ref.~\cite{Dominguez:2000xa} showed that the strength
of such terms is consistent with zero.\footnote{
They did this by fitting a $d\!=\!2$ finite energy sum rule,
which is apparently well satisfied at the used relevant scales
\cite{Maltman:1998rh,Dominguez:1999wy}, 
to the new ALEPH data on spectral functions extracted 
from the $\tau$ decay measurements.}

When subtracting the quark mass contributions
(\ref{drtmud1})--(\ref{drtmud2}) from Eq.~(\ref{rtgen}), we
end up with the massless QCD observable
\begin{eqnarray}
r_{\tau} &\equiv& 
r_{\tau}(\triangle S\!=\!0; m_{u,d} =0) =
r_{\tau}(\triangle S\!=\!0) -
\delta r_{\tau}(\triangle S\!=\!0)_{m_{u,d}\not=0}
\label{rtmud1=0}
\\
& = & - \frac{ {\rm i}}{2 \pi} \int_{|s|=m^2_{\tau}}
\frac{ds}{s} \left( 1\!-\!\frac{s}{m^2_{\tau}} \right)^3
\left( 1 + \frac{s}{m^2_{\tau}} \right) D(-s) \ ,
\label{rtmud2=0}
\\
&=& \frac{1}{2 \pi} \int_{-\pi}^{\pi} dy \left( 1 + e^{{\rm i} y} \right)^3
\left( 1 - e^{{\rm i} y} \right) 
D(-s\!=\!m_{\tau}^2 e^{{\rm i}y} ) \ ,
\label{rtmud3=0}
\end{eqnarray}
with the
canonically normalized massless Adler function $D(Q^2\!\equiv\!-s)$
defined in Eq.~(\ref{Dexpan}).

The perturbation coefficients $d_j = d_j(\mu^2/Q^2; c_2,\ldots, c_j)$ 
in Eq.~(\ref{Dexpan})
depend on the RScl $\mu^2$ and the RSch parameters $c_j$ in 
a known specific manner, because $D(Q^2)$ is RScl-- and RSch--independent. 
Knowing them at a specific RScl and 
RSch, we know them at any RScl and RSch -- 
see Eqs.~(\ref{d1xi2})--(\ref{d4xi2}) in the Appendix.
The first two coefficients have already been calculated,
and in the ${\overline {\rm MS}}$ RSch and at RScl $\mu^2\!=\!Q^2$,
for $n_f\!=\!3$, they are: $d_1^{(0)} = 1.6398$ \cite{coeffs1};
$d_2^{(0)} = 6.3710$ \cite{coeffs2} (the superscript (0)
denotes the value at the aforementioned RScl and RSch). 

In order to extract the experimental value of $r_{\tau}$ 
(\ref{rtmud1=0}),
according to Eq.~(\ref{rtgen0})
we need to use the values 
$R_{\tau}(\triangle S\!=\!0)$ of Eq.~(\ref{Rtauexp}), 
$\delta_{\rm EW} = 0.0194 \pm 0.0050$ 
\cite{Braaten:1990ef,Marciano:1988vm,Braaten:1992qm},
$\delta^{\prime}_{\rm EW} = 0.0010$ \cite{Braaten:1990ef},
$\delta r_{\tau}(\triangle S\!\not=\!0)_{m_{u,d}\not=0}$ 
(\ref{drtmud2}),
and in addition the value of the CKM matrix element $|V_{ud}|$,
which we take as\footnote{
The standard model (SM) unitarity--constrained fit predicts
$|V_{ud}|\!=\!0.9749 \pm 0.0008$ \cite{Groom:2000in}.
However, the values extracted from the
decays of mirror nuclei are lower:
$|V_{ud}| = 0.9740 \pm 0.0010$. This extraction
has significant theoretical uncertainties
(see \cite{Groom:2000in} for further References).
The values extracted from
neutron decays are even lower 
$|V_{ud}| = 0.9728 \pm 0.0012$
(\cite{Groom:2000in} and References therein),
but appear to have smaller theoretical uncertainties.
In view of all these considerations, we take in our analysis
the value range as given in Eq.~(\ref{Vud}).
Here, the central value is from the unitarity--constrained fit, 
and the uncertainty is increased so that (\ref{Vud})
covers all the values from the decays of mirror nuclei
and the upper half of the interval from neutron decays.}
\begin{equation}
|V_{ud}| = 0.9749 \pm 0.0021 \ .
\label{Vud}
\end{equation}
This leads us, via Eqs.~(\ref{Rtauexp}), 
(\ref{rtgen0}) and (\ref{drtmud2}), 
to the following values for the massless QCD observable (\ref{rtmud1=0})
\begin{eqnarray}
r_{\tau} &\equiv& r_{\tau}(\triangle S\!=\!0; m_{u,d}\!=\!0)
\nonumber\\
&=& 0.1960 \pm 0.0059_{\rm exp.} \pm 0.0059_{\rm EW} \pm 0.0051_{\rm CKM}
\label{rtauexp}
\\
&=& 0.1960 \pm 0.0098 \ .
\label{rtauexps}
\end{eqnarray}
In Eq.~(\ref{rtauexps}), the
three uncertainties of Eq.~(\ref{rtauexp}) were added in quadrature.

The values (\ref{rtauexp}), together with the
contour integral expression (\ref{rtmud3=0}),
will be the starting
point for our massless QCD resummation analyses of hadronic
$\tau$ decays.

\section{Modified Borel transforms}
\label{sec:modBT}

The most straightforward way to perform the
resummation for $r_{\tau}$ given in
Eqs.~(\ref{rtmud2=0})--(\ref{rtmud3=0}) 
would be to insert the known
truncated perturbation series (TPS) for $D(-s)$ given in
Eq.~(\ref{Dexpan}) and perform the
momentum--contour integration, i.e., the approach of
\cite{Pivovarov:1992rh}. The method is, firstly, fraught with
ambiguities from the choice of RScl and RSch.
The final result contains residual, but significant,
RScl and RSch dependence, due to the truncation of the
series in Eq.~(\ref{Dexpan}). Secondly, the method does not
incorporate the known renormalon structure of the
Adler function's Borel transform ${\widetilde D}(b)$.

In a previous paper \cite{Cvetic:2001sn}, two of us addressed the second
problem, by employing in the resummation a (ordinary) Borel 
transform ${\widetilde D}(b)$ that includes the
leading IR renormalon via the ansatz
${\widetilde D}(b) = R(b)/(1 - b/2)^{1+\nu}$,
and by introducing in addition
conformal transformations $b=b(w)$ in order to map
sufficiently far away from the origin the singularities of the
UV (and the remaining IR) renormalons. 
However, the integrand in the (ordinary)
Borel integral is RScl-- and RSch--dependent. This can
be inferred from the definition
\begin{equation}
D(Q^2)= \frac{1}{\beta_0}
\int^\infty_0 db 
\exp\left[ - \frac{b}{{\beta_0} a(\mu^2; c_2, \ldots)} \right] 
\widetilde D(b; \mu^2/Q^2, c_2, \ldots) \ ,
\label{Borel-integral}
\end{equation}
and from the expansion of the transform around the origin
\begin{equation}
\widetilde D(b; \mu^2/Q^2, c_2, \ldots)= 1 + \frac{d_1(\mu^2/Q^2)}{1!}
\left(\frac{b}{\beta_0}\right) +
\sum_{n=2}^\infty \frac{d_n(\mu^2/Q^2; c_2, \ldots, c_n)}{n!} 
\left(\frac{b}{\beta_0}\right)^n \ .
\label{SBexp}
\end{equation}
For example, direct application of the derivative with respect
to the RScl $\mu^2$ gives
\begin{eqnarray}
\frac{\partial}{\beta_0 \partial \ln \mu^2} \left[ 
{\rm e}^{-b/(\beta_0 a)} \widetilde D(b) \right] &=&
{\rm e}^{-b/(\beta_0 a)} {\Bigg \{} \left[
\frac{ \beta(a) }{ \beta_0 a^2 } + 1 \right] \frac{b}{\beta_0} +
\left[ d_1 \left( \frac{ \beta(a) }{ \beta_0 a^2 } + 1 \right)
+ \frac{c_1}{2} \right] \frac{1}{1!} 
\left( \frac{b}{\beta_0} \right)^2 
\nonumber\\
&&
+ \left[ d_2 \left( \frac{ \beta(a) }{ \beta_0 a^2 } + 1 \right)
+ \frac{1}{3} ( 2 c_1 d_1 + c_2) \right] \frac{1}{2!}
\left( \frac{b}{\beta_0} \right)^3 + \cdots {\Bigg \}} \ ,
\label{SBder}
\end{eqnarray}
using the notations of Eq.~(\ref{aRGE}). Once going
beyond the one--loop approximation of the 
RGE--evolution (\ref{aRGE}), the integrand is RScl--dependent.
If we knew the exact expression of the integrand, the
total integral (\ref{Borel-integral}) would be
RScl--independent. However, since we have a TPS available 
for ${\widetilde D}(b)$, we are forced to use
a TPS for $R(b)\!\equiv\!(1 - b/2)^{1+\nu} {\widetilde D}(b)$. 
This truncation then results in the residual
RScl and RSch dependence of the resummed result for 
$D(Q^2)$ and thus for $r_{\tau}$.
In \cite{Cvetic:2001sn} we fixed the
RScl--parameter $\xi^2\equiv\mu^2/Q^2$ according to the
principle of minimal sensitivity (PMS), i.e.,
$\partial r_{\tau}/\partial \xi^2 = 0$. The RSch was chosen to
be ${\overline {\rm MS}}$.

In the present paper
we apply, instead of the ordinary Borel transform,
several variants of the modified Borel transform
${\overline {D}}(b)$ of the Adler function $D(Q^2)$. 
The integrand will be RScl--independent.
This Borel transform was
introduced for QCD and QED (quasi)observables by Grunberg
\cite{Grunberg:1993hf}, who in turn constructed them on the basis
of the modified Borel transformations of Ref.~\cite{modBT}.
The integral transformation for ${\overline {D}}(b)$ is
written in the form
\begin{equation}
D(Q^2) = \frac{1}{\beta_0} \int_0^\infty db \exp
\left[ - \frac{ (\rho_1(Q^2)\!+\!{\tilde c}) b }{ \beta_0} \right]
{\overline {D}}(b; {\tilde c}) \ .
\label{invBT}
\end{equation}
Here, ${\overline {D}}(b; {\tilde c}) = 
\exp({\tilde c}\;b/\beta_0) {\overline {D}}(b;0)$ 
and has no $Q^2$--dependence;
${\tilde c}$ is a specific arbitrary constant;
$\rho_1$ is the first RScl and RSch invariant
\cite{PMS} of the Adler function
\begin{equation}
\rho_1(Q^2) = - d_1(\mu^2/Q^2) + 
\beta_0 \ln \frac{ \mu^2 }{ {\widetilde \Lambda}^2 } =
\beta_0 \ln \frac{ Q^2 }{ {\overline \Lambda}^2 } \ ,
\label{rho1}
\end{equation}
where ${\widetilde {\Lambda}}$ is the universal scale appearing
in the Stevenson's equation \cite{PMS}, and
${\overline {\Lambda}}$ is a scale which depends on the
observable but is RScl-- and RSch--independent.
We note that $\rho_1(Q^2) = [ 1/a^{{\rm (1-loop)}}(Q^2) + c]$,
where $c$ is a constant.
Therefore, ${\overline D}(b; {\tilde c})$ reduces to the
ordinary Borel transform ${\widetilde D}(b; \mu^2/Q^2\!=\!1, \ldots)$
times the factor $\exp[(c\!+\!{\tilde c}) b/\beta_0]$, when higher 
than one--loop effects are ignored (large--$\beta_0$ approximation).

Since $D(Q^2)$ and $\rho_1(Q^2)$ in Eq.~(\ref{invBT}) 
are RScl-- and RSch--independent, so
${\overline {D}}(b; {\tilde c})$ possesses no explicit
RScl-- and RSch--dependence. 
However, as shown in the Appendix,
the modified Borel transformation (\ref{invBT})
is one in a large class of Borel transformations, where 
each of them corresponds to a specific choice of the RScl and
RSch. The transformation (\ref{invBT}) is obtained
from this class by choosing the so called modified
`t Hooft RSch (mtH RSch: $c_j\!=\!c_1^j$, $j\!=\!2,3,\ldots$).
We also show that in this case the constant ${\tilde c}$
is in fact the RScl parameter ${\tilde c}\!=\!\beta_0 \ln(\xi^2)$, 
where $\xi^2\!\equiv\!\mu^2/Q^2$.
This special Borel transformation has the remarkable
property that the integrand in the Borel integral is
RScl--independent (${\tilde c}$--independent).
Therefore, we can call it RScl--independent
Borel transformation, since the change of the RScl
only changes the convention of separating the
(RScl--invariant) integrand in Eq.~(\ref{invBT})
into two factors. 

Yet another  useful property of the modified
Borel transform (\ref{invBT}) is the following:
due to the very simple $Q^2$--dependence of the integrand 
in Eq.~(\ref{invBT}), the contour integration
(\ref{rtmud3=0}) in the complex momentum plane
can be performed analytically, leading to a
rather simple expression for the observable $r_{\tau}$,
as will be shown in the next section. This is not
possible if the mtH RSch is abandoned. 

The coefficients of expansion of ${\overline {D}}(b;{\tilde c})$
in Eq.~(\ref{invBT}) around $b\!=\!0$ 
are related with those of the expansion of $D(Q^2)$ 
most easily when we use for the latter expansion the specific 
mtH RSch $c_k = c_1^k$ 
($k \geq 2$) and RScl $\mu^2\!=\!Q^2$ \cite{Grunberg:1993hf}
\begin{eqnarray}
{\overline D}(b;{\tilde c}) &=& 
\frac{ (c_1 b/\beta_0)^{c_1 b/\beta_0} }{ \Gamma(1 + c_1 b/\beta_0) } 
\exp \left( \frac{ ({\tilde c}\!-\!{\tilde d}_1) b }{\beta_0} \right)
{\Bigg \{} 1 + \frac{({\tilde d}_1\!-\!c_1)}{ (1\!+\!c_1 b/\beta_0)} 
\left( \frac{b}{\beta_0} \right)
\nonumber\\
&& + \sum_{n=2}^{\infty} \frac{ ( {\tilde d}_n\!-\!c_1 {\tilde d}_{n-1} ) }
{(1\!+\!c_1 b/\beta_0) (2\!+\!c_1 b/\beta_0) \cdots (n\!+\!c_1 b/\beta_0) } 
\left( \frac{b}{\beta_0} \right)^n {\Bigg \}} \ . 
\label{barDexp}
\end{eqnarray}
Here, ${\tilde d}_j = d_j(\mu^2/Q^2; c_2, \ldots, c_j)$
with $\mu^2\!=\!Q^2$, $c_k\!=\!c_1^k$ ($k\!=\!2,\ldots,j$).
We refer to the Appendix for details.
Expanding each term on the right--hand side in powers of $b$
then yields the expansion of ${\overline D}(b;{\tilde c})$ 
around $b\!=\!0$.
Here we can see again that in the large--$\beta_0$
approximation ($c_1 \to +0$) the above expansion reduces to 
the expansion (\ref{SBexp}) of the ordinary Borel transform 
${\widetilde D}(b; 1, \ldots)$ times the factor
$\exp[({\tilde c}\!-\!{\tilde d}_1) b/\beta_0 ]$.


The ordinary Borel transform ${\widetilde D}(b)$ is known
to have singularities  at $b\!=\!-1, -2, \ldots$
(UV renormalons) and at $b\!=\!2,3,\ldots$ (IR renormalons).
The renormalon resummation of $D(Q^2)$ and of the
hadronic $\tau$ decay width in the large--$\beta_0$
limit has been performed in 
Refs.~\cite{BBB,Neubert:1996gd,Lovett-Turner:1995ti}.
The IR renormalons on the contour of the
Borel integration (\ref{Borel-integral}) cause ambiguities,
above all the leading IR renormalon ($b\!=\!2$).
The singularity of ${\widetilde D}(b)$ at $b\!=\!n$ has the
form $1/(1-b/n)^{2+\gamma^{\prime}_n+n c_1/\beta_0}$
for $n \geq 3$, and $1/(1-b/2)^{1+\gamma^{\prime}_2 + 2 c_1/\beta_0}$
at $b\!=\!2$, where $\gamma^{\prime}_n$ is the
one--loop anomalous dimension of the operators
corresponding to the $1/Q^{2 n}$--terms in the operator
product expansion (OPE) of the Adler function $D(Q^2)$.
In the case $n\!=\!2$, it is known that $\gamma^{\prime}\!=\!0$
\cite{Mueller:1985vh}. Thus, at the leading IR renormalon
we have ${\widetilde D}(b) \propto 1/(1-b/2)^{1 + \nu}$,
with $\nu=2 c_1/\beta_0$ ($=\!1.580$, when $n_f\!=\!3$).

On the other hand, the modified Borel transform ${\overline D}(b)$
in Eq.~(\ref{invBT}) has the (IR and UV) 
renormalon singularities at the same locations
as ${\widetilde D}(b)$, but with simpler powers \cite{Grunberg:1993hf,modBT}
-- the IR renormalon singularities are of the form 
$1/(1-b/n)^{2 + \gamma^{\prime}_n}$
for $n \geq 3$, and $1/(1-b/2)^1$ for the leading IR singularity\footnote{
This is true even when the mtH RSch is abandoned and the
integral transformation (cf.~Appendix) becomes considerably more
complicated -- as follows from the considerations of 
Ref.~\cite{modBT}.}
[see Eq.~(\ref{sing2}) in the Appendix, with $\kappa\!=\!n$].
Therefore, we can define a new function ${\overline R}(b;{\tilde c})$
\begin{equation}
{\overline R}(b;{\tilde c}) \equiv 
(1 - b/2) {\overline D}(b; {\tilde c}) \ ,
\label{barR}
\end{equation} 
which has a considerably softened singularity at $b\!=\!2$
(cut instead of pole). 
This function can be determined 
by resummations of the TPS of ${\overline R}(b;{\tilde c})$
(cf.~also Ref.~\cite{Lee:1997yk}).
The latter TPS is known to the same order as 
${\overline D}(b; {\tilde c})$ -- see Eq.~(\ref{barDexp}):
$d_1^{(0)}$ and $d_2^{(0)}$ 
($\Leftrightarrow {\tilde d}_1, {\tilde d}_2$)
are known exactly, and $d_3^{(0)}$ ($\Leftrightarrow {\tilde d}_3$)
is known approximately. 
The same is true for the
modified functions ${\overline R}_j(b;{\tilde c})$
($j\!=\!1,2,3$) discussed below, Eqs.~(\ref{barR1})--(\ref{barR3}).

The function ${\overline D}(b;{\tilde c})$ as defined in 
Eq.~(\ref{invBT})
has additional singularities, as seen from the expansion (\ref{barDexp}): 
1.) it is nonanalytic at $b\!=\!0$, due to the (finite) 
factor $(c_1 b/\beta_0)^{c_1 b/\beta_0}$;
2.) it has additional (spurious) poles at $b\!=\!-\beta_0/c_1$
($\approx -1.266$ when $n_f\!=\!3$), $b\!=\!-2 \beta_0/c_1$, etc.
In approximate numerical evaluations of $D(Q^2)$ and $r_{\tau}$
via ${\overline D}(b;{\tilde c})$, such singularities may have disturbing
effects unless they are sufficiently far away from the origin. Therefore,
we can define the following variants which will be used
in (approximate) resummations
\begin{eqnarray}
{\overline R}_1(b;{\tilde c}) & = &  (1 - b/2 ) 
(c_1 b/\beta_0)^{-c_1 b/\beta_0} \Gamma(1 + c_1 b/\beta_0)
{\rm e}^{c_1 b/\beta_0} {\overline D}(b;{\tilde c}) 
\label{barR1}
\\
&=& (1\!-\!b/2) \exp [ ( {\tilde c}\!+\!c_1\!-\!{\tilde d}_1) b/\beta_0 ]
{\Bigg \{} 1 + \frac{({\tilde d}_1\!-\!c_1)}{ (1\!+\!c_1 b/\beta_0)} 
\left( \frac{b}{\beta_0} \right) + \ldots {\Bigg \}} \ ,
\label{barR1b}
\\
{\overline R}_2(b;{\tilde c}) & = & {\overline R}_1(b;{\tilde c}) 
\frac{(1 + c_1 b/\beta_0)}{(2 + c_1 b/\beta_0)} \ ,
\label{barR2}
\\
{\overline R}_3(b;{\tilde c}) & = & {\overline R}_1(b;{\tilde c}) 
\frac{(1 + c_1 b/\beta_0)(2 + c_1 b/\beta_0)}
{(3 + c_1 b/\beta_0)^2} \ .
\label{barR3}
\end{eqnarray}
We recall that ${\overline R}_j(b;{\tilde c}) = 
\exp({\tilde c}\;b/\beta_0) {\overline R}_j(b;0)$.
In the above expressions, we restored analyticity at $b\!=\!0$
by factoring out, instead of the factor $(c_1 b/\beta_0)^{c_1 b/\beta_0}$,
the combination 
$(c_1 b/\beta_0)^{c_1 b/\beta_0}
{\rm e}^{-c_1 b/\beta_0}/\Gamma(1 + c_1 b/\beta_0)$. 
Our main motivation for this lies in the following:
the factor 
$(c_1 b/\beta_0)^{c_1 b/\beta_0}$ is increasing extremely fast
with increasing $b$, while the aforementioned combination
has weak $b$--dependence
\begin{equation}
\frac{(c_1 b/\beta_0)^{c_1 b/\beta_0} {\rm e}^{-c_1 b/\beta_0}}
{\Gamma(1 + c_1 b/\beta_0)} =
\frac{1}{\sqrt{2 \pi c_1 b/\beta_0}} \quad {\rm when} \quad
b \to \infty \ .
\label{factor}
\end{equation}
Hence, functions $R_j(b;{\tilde c})$ behave at large $b$ 
roughly as $R(b;{\tilde c})$, or ${\overline D}(b;{\tilde c})$, 
\begin{equation}
{\overline R}_j(b;{\tilde c}) \sim (1 - b/2) \sqrt{2 \pi c_1 b/\beta_0}
\;{\overline D}(b;{\tilde c}) \qquad {\rm when} \ b \to \infty \ ,
\label{asympt}
\end{equation}
i.e., they neither decrease nor increase
violently. Therefore, any approximate
resummation method will have better chance when applied
to them than to an extremely fast increasing or decreasing version.
On the other hand, if we just factored out the
factor $(c_1 b/\beta_0)^{c_1 b/\beta_0}$, the resulting
function, though analytic at $b\!=\!0$, would at large $b$
decrease violently, as 
$\sim\!(c_1 b/\beta_0)^{- c_1 b/\beta_0} {\overline D}(b;{\tilde c})$.

The function $R_1(b;{\tilde c})$ has spurious (unphysical) poles
at $b\!=\!-\beta_0/c_1$, $-2 \beta_0/c_1$, $\ldots$ 
($\!\approx -1.261$, $-2.53$, when $n_f\!=\!3$).
But the function $R_2$ has a possible advantage over $R_1$
in resummations, since it has no spurious (nonphysical)
pole at $b\!=\!-\beta_0/c_1$,
and $R_3$ has no such poles at $b\!=\!-\beta_0/c_1$
and $-2\beta_0/c_1$.

Since the coefficients $d_1^{(0)}$ and $d_2^{(0)}$ 
($ \Rightarrow {\tilde d}_1, {\tilde d}_2$) of the (massless) Adler 
function are known, the power expansion of 
${\overline R}_j(b;{\tilde c})$
is known to the next-to-next-to-leading order 
(NNLO, including the term $\sim\!b^2$) via
Eqs.~(\ref{barDexp})--(\ref{barR3}). 
In a previous work \cite{Cvetic:2001sn}, we presented an argument,
via a bilocal expansion of the Borel amplitude 
${\widetilde D}(b)$, that $d_3^{(0)} \approx 25 \pm 5$.   
We also discussed there the estimates of $d_3^{(0)}$
presented by other authors, and concluded that the following
estimate is rather safe
\begin{equation}
d_3^{(0)} 
\left[ \equiv d_3(\mu^2\!=\!Q^2; {\overline {\rm MS}}) \right]
= 25 \pm 10 \ .
\label{d3est}
\end{equation}
We will use the above values which allow us to obtain
the power series of ${\overline R}_j(b;{\tilde c})$ 
up to ${\rm N}^3 {\rm LO}$ ($\sim\!b^3$).

\section{Resummation procedure}
\label{sec:resummation}

We will apply summations to the ${\rm N}^3 {\rm LO}$
truncated power series (TPS) of the 
functions ${\overline R}_j(b;{\tilde c})$.
However, in order to obtain the (resummed) values of the
massless QCD observable $r_{\tau}$ (\ref{rtmud1=0}),
we have to perform first the complex momentum contour integration
(\ref{rtmud3=0}), with the massless Adler function there
having the integral form (\ref{invBT}) in terms of the
invariant Borel transform 
${\overline D}(b; Q^2\!=\!m^2_{\tau} e^{{\rm i} y})$,
i.e., in terms of the related functions
${\overline R}_j$ (\ref{barR1})--(\ref{barR3}).
This angular $y$--integration can be performed exactly, because the
$y$--dependence of the invariant $\rho_1$ (\ref{rho1})
appearing in Eq.~(\ref{invBT}) is simple
\begin{equation}
\rho_1(m^2_{\tau} e^{{\rm i} y} ) = \rho_1(m^2_{\tau}) + 
{\rm i} \beta_0 y  \quad \Rightarrow
\label{rho1y}
\end{equation}
\begin{eqnarray}
r_{\tau} &=& \frac{1}{\pi \beta_0} \text{Re}\left[ 
\int_{0-i\varepsilon}^{\infty-i\varepsilon} db
\exp \left[ 
- \frac{ \left( {\rho}_1(m^2_{\tau})\!+\!{\tilde c} \right) b}
{\beta_0} \right]
\frac{ \sin( \pi b) }{ b (1\!-\!b) (1\!-\!b/3) (1\!-\!b/4) }
{\overline D}(b; {\tilde c}) \right] \ ,
\label{rtinvBT}
\\
& = & \frac{1}{\pi \beta_0} \text{Re} {\Bigg \{}
\int_{0-i\varepsilon}^{\infty-i\varepsilon} db
\exp \left[ 
- \frac{ \left({\rho}_1(m^2_{\tau};a_0)\!+\!{\tilde c} \right) b}
{\beta_0} \right] 
\frac{ \sin( \pi b) }{ b (1\!-\!b) (1\!-\!b/2)(1\!-\!b/3) (1\!-\!b/4) }
\nonumber\\
&& \times
\frac{(c_1 b/\beta_0)^{c_1 b/\beta_0} e^{- c_1 b/\beta_0} }
{\Gamma(1\!+\!c_1 b/\beta_0)} f_j(b) {\overline R}_j(b;{\tilde c}) 
{\Bigg \}} \ ,
\label{rtRj}
\end{eqnarray}
where 
\begin{eqnarray}
f_j(b) & = & \left \{
\begin{array}{l l} 
1 & \text{if $j=1$} \ , \\
(2\!+\!c_1 b/\beta_0)/(1\!+\!c_1 b/\beta_0) & \text{if $j=2$} \ , \\
(3\!+\!c_1 b/\beta_0)^2/(1\!+\!c_1 b/\beta_0)/(2\!+\!c_1 b/\beta_0) & 
\text{if $j=3$} \ .
\end{array}
\right \}
\label{fjs}
\end{eqnarray}
The integration contour in Eqs.~(\ref{rtinvBT})--(\ref{rtRj}) is chosen
slightly below (or above) the positive axis, in order to
avoid possible singularities of the integrand on the positive axis.
When knowing reasonably well the RScl-- and RSch--invariant
functions $R_j(b)$ in Eq.~(\ref{rtRj}), e.g. via resummations as
discussed below, the massless QCD observable $r_{\tau}$
becomes an expression whose value depends uniquely on
the value of the QCD coupling parameter, e.g. on
$\alpha_s(m^2_{\tau}; {\overline {\rm MS}})\!\equiv\!\pi a_0$.
This dependence originates from the known dependence 
of the invariant $\rho_1(m^2_{\tau})$
in the exponent in Eq.~(\ref{rtRj}), the latter being determined
by an integrated version of the RGE, called also
the (unsubtracted) Stevenson equation \cite{PMS}
\begin{eqnarray}
\rho_1(m^2_{\tau}; a_0) & \equiv & 
- d_1^{(0)} + \beta_0 \ln \frac{m^2_{\tau}}{{\widetilde {\Lambda}}}
\nonumber\\
&=& - d_1^{(0)} + \frac{1}{a_0} + c_1 \ln \left(
\frac{ c_1 a_0}{1\!+\!c_1 a_0} \right) + \int_0^{a_0} dx \left[
\frac{1}{x^2(1\!+\!c_1 x)} + \frac{{\beta}_0}
{{\beta}_{\overline {\rm MS}}(x)}  \right] \ ,
\label{Steve}
\end{eqnarray}
where the perturbation expansion of the last term is
given via the definition (\ref{aRGE}) of the $\beta$--function
\begin{equation}
{\beta}_{\overline {\rm MS}}(x)/\beta_0 =
- x^2 (1\!+\!c_1 x\!+\!c_2^{\overline {\rm MS}} x^2\!+\!
c_3^{\overline {\rm MS}} x^3\!+\!\cdots ) \ .
\label{betaMS}
\end{equation}
Here, the ${\overline {\rm MS}}$ coefficients are functions of the
number of active quark flavors $n_f$ and are known up to
the ${\rm N}^3 {\rm LO}$ ($c_3^{\overline {\rm MS}}$, four--loop) 
\cite{vanRitbergen:1997va}.
The number of active quark flavors is assumed to be here $n_f\!=\!3$,
because the scale of the process is $\sqrt{ |Q^2|} =\!m_{\tau}$ 
($\approx\!1.777$ GeV).

We can note from Eq.~(\ref{rtRj}) that ambiguity of 
integration over $b$ in $r_{\tau}$ at the first infrared
renormalon singularity of ${\overline D}(b)$ at $b\!=\!2$
is suppressed,
because the factor $\sin (\pi b)$ has a zero there.
This is similar to the one--loop approximation in the
approach with the ordinary Borel transform
\cite{Caprini:1999wg,Altarelli:1995vz}. 
However, here the absence of the ambiguity
is not due to an approximation, it is exact and due to the
discussed RScl--invariant Borel approach.

Although the ambiguity at $b\!=\!2$
is suppressed in the Borel--type of integration (\ref{rtRj}),
we wish to emphasize that it is nonetheless very important to
factor out the leading renormalon factor $1/(1\!-\!b/2)$ there
[according to Eq.~(\ref{barR})]. This is so because any resummation
of a ${\rm TPS}(b)$ represents also a quasianalytic continuation  
of the corresponding function into the region away from the
origin, and such continuation is of a better quality when
there are as few singularities near the origin as possible.
The functions which we will resum in Eq.~(\ref{rtRj}) are
${\overline R}_j(b;{\tilde c})$, i.e.,
the functions which have their original pole singularity at $b\!=\!2$
factored away according to Eq.~(\ref{barR}). The remaining
singularity in ${\overline R}_j(b;{\tilde c})$ is then 
significantly weaker, it is
a cut of the type $\sim\!\ln(1\!-\!b/2)$ [see~Eq.~(\ref{sing2})
with $\alpha\!=\!1$ and $\kappa\!=\!2$, in conjunction with
Eq.~(\ref{barDcalF})].
In this way, in the relatively wide region $b \stackrel{<}{\sim} 2$
we can achieve reasonably good values of the
integrand of Eq.~(\ref{rtRj}), which leads then
to good predictions for $r_{\tau}$ via Eq.~(\ref{rtRj})
as functions of $\alpha_s^{\overline {\rm MS}}(m^2_{\tau})$.
If we did not factor out the mentioned singularity,
the obtained (resummed) values would definitely lead to
bad values of the integrand at $b\!\sim\!2$, the
fact which would influence the predicted values of $r_{\tau}$
and thus of $\alpha_s^{\overline {\rm MS}}(m^2_{\tau})$.
The numerical importance of factoring out the leading IR
renormalon singularity of the Borel transforms of the Adler function
at $b\!=\!2$ before doing resummation was pointed out
in our previous work \cite{Cvetic:2001sn}, 
which involved ordinary Borel transforms.
There we showed that the predicted values of 
$\alpha_s^{\overline {\rm MS}}(m^2_{\tau})$
from $r_{\tau}$ depend crucially on whether
this factoring--out procedure has been performed.

The functions ${\overline R}_j(b; {\tilde c})$ in Eq.~(\ref{rtRj}), 
whose TPS's we want to resum via methods of quasianalytic continuation, 
have the singularities near the origin which are negative
(at $b=-1$ and lower). The nearest singularity on the positive
axis is at $b=3$. At $b\!=\!2$ (${\rm IR}_2$) there is a weaker
logarithmic singularity. The negative singularities near the origin
constrain the convergence radius of the perturbation
(power) series of ${\overline R}_j(b; {\tilde c})$'s to $r\!=\!1$,
and thus represent a possible hindering element to efficient
resummations. In our previous work \cite{Cvetic:2001sn}
we proposed how to
extend the convergence radius up to ${\rm IR}_2$ -- by either
of the following two conformal transformations $w\!=\!w(b)$:
\begin{equation}
w_3(b) = \frac{\sqrt{1+b}-\sqrt{1-b/3}}{\sqrt{1+b}+\sqrt{1-b/3}} \ ,
\qquad
w_4(b)  = \frac{\sqrt{1+b}-\sqrt{1-b/4}}{\sqrt{1+b}+\sqrt{1-b/4}} \ .
\label{ct34}
\end{equation}
Conformal transformation $w_3(b)$ maps all the renormalon
singularities to the unit circle in the $w$--plane, except
for the first IR renormalon which is mapped to $w_3(b\!=\!2) = 1/2$.
We further note that $w_3(b\!=\!3) = 1$, and $w_3(b\!=\!-1) = -1$.
Furthermore, all the spurious singularities $b\!=\!- n \beta_0/c_1$
($< - 1.265$) are also sent to the unit circle. Conformal transformation
$w_4(b)$ does the same thing, except that the first
two IR renormalon singularities are both inside the unit
circle: $w_4(b\!=\!2) \approx 0.42$, $w_4(b\!=\!3) = 0.6$.
We further note that $w_4(b\!=\!4) = 1$, and $w_4(b\!=\!-1) = -1$.
The inverse transformations are
\begin{equation}
b(w_3) = \frac{3 w_3}{(1 - w_3 + w_3^2)} \ ,
\qquad
b(w_4) =  \frac{(16/5) w_4}{(1 - (6/5) w_4 + w^2_4)} \ ,
\label{ct34inv}
\end{equation}
which are monotonously increasing functions for $0 < w_3, w_4 < 1$.
We can now reexpress the (${\rm N}^3 {\rm LO}$) TPS of 
${\overline R}_j(b;{\tilde c})$ (in powers of $b$)
as (${\rm N}^3 {\rm LO}$) TPS in powers of
$w_3$ or $w_4$, by simply using the power expansions of
Eq.~(\ref{ct34inv}).
The advantage of using this form of TPS for resummations
lies in the fact that the convergence radius for the
power series of ${\overline R}_j$ is now $r_{w_3}\!=\!1/2$
in the $w_3$--plane, and $r_{w_4}\!\approx\!0.42$ in the
$w_4$--plane, and the circle of convergence reaches thus
the first IR renormalon singularity $b(w_k)\!=\!2$ ($k\!=\!3,4$), 
in contrast to the case of nontransformed $b$. In this way,
the hindering influence of the UV singularities (negative $b$'s)
has been significantly weakened. The mapping $w_4(b)$
apparently suppresses even more strongly the influence of the
UV singularities than $w_3(b)$, but probably less strongly the 
influence of the NLO IR renormalon singularity ($b\!=\!3$).
The final formula for $r_{\tau}$ in this formulation follows
directly from the form (\ref{rtRj})
\begin{eqnarray}
r_{\tau} &=& 
\frac{1}{\pi \beta_0} {\rm Re} {\Bigg \{} e^{ - \rm i \phi}
\int_0^{1} d x \frac{d b(w)}{d w}
\exp \left[ 
- \frac{ \left({\rho}_1(m^2_{\tau};a_0)\!+\!{\tilde c} \right)
b(w)}{\beta_0} \right] 
\nonumber\\
&& \times 
\frac{ \sin \left( \pi b(w) \right) }
{ b(w) \left( 1\!-\!b(w) \right) \left( 1\!-\!b(w)/2 \right)
\left( 1\!-\!b(w)/3 \right)  \left( 1\!-\!b(w)/4 \right) }
\nonumber\\
&& \times
\frac{\left( c_1 b(w)/\beta_0 \right)^{c_1 b(w)/\beta_0} 
e^{ - c_1 b(w)/\beta_0} }
{ \Gamma \left( 1\!+\!c_1 b(w)/\beta_0 \right) } f_j(b(w)) 
{\overline R}_j(b(w); {\tilde c}){\Bigg |}_{w=x e^{\rm - i \phi}} 
{\Bigg \}}\ ,
\label{rtRjw}
\end{eqnarray}
where $f_j(b)$'s are fiven by Eq.~(\ref{fjs}),
$w$ stands for $w_3$ or $w_4$, and we integrate actually only
up to a $w_{\rm max}$ corresponding to $b \approx 4$
(for $w_3$, $\phi_3\!=\!0.50536$,
$w_{3 {\rm max}}\!=\!\exp(-{\rm i} \phi_3)$; 
for $w_4$, $\phi_4\!=\!0.1 \approx 0$,
$w_{4 {\rm max}}\!=\!\exp(-{\rm i} \phi_4) \approx 1$).
This upper bound on $b$ is justified because the contributions
from higher $b$'s are very strongly suppressed due to the
exponent in Eq.~(\ref{rtRjw}). Strictly speaking, the 
path in the $w_3$-plane should be along the positive
axis (below it) from $w_3\!=\!0$ to $w_3\!=\!1$, and then along 
the arc (inner side) of the unit circle between 
$w_3\!=\!1$ and $w_3\!=\!w_3(b\!=\!4)\!=\!\exp(-{\rm i} \phi_3)$, 
as shown in Fig.~\ref{fig1}.
However, for practical calculations,
it is much more convenient to use the integration along the
ray $w_3\!=\!x \exp( - {\rm i} \phi_3)$ ($0 \leq x \leq 1$),
as denoted in Eq.~(\ref{rtRjw}) and shown in Fig.~\ref{fig1}.
Both paths give the same answer, since the closed contour
in Fig.~\ref{fig1} does not contain any singularities of
the integrand. In the case of $w_4$, the trick is the same,
and we choose $\phi_4\!=\!0.1$ instead of $\phi_4\!=\!0$
for the ray (see Fig.~\ref{fig2}), 
in order to avoid any possible problems 
with numerical instability that would
otherwise arise from the too extreme vicinity of the integration path
to the possible singularities of the integrand.

At this point, we can thus already regard the ${\rm N}^3 {\rm LO}$ TPS
of the function
${\overline R}_j(b(w_k);{\tilde c})\!\equiv\!\exp
[{\tilde c}\;b(w_k)/\beta_0] {\overline R}_j(b(w_k); 0)$ 
($k\!=\!3,4$) as a form of resummation
of ${\overline R}_j$, solely via the mappings (\ref{ct34}).
Consequently, expressions (\ref{rtRjw}) evaluated using the
aforementioned ${\rm N}^3 {\rm LO}$ TPS of 
${\overline R}_j(b(w_k); {\tilde c})$
can be regarded as our resummed predictions of 
$r_{\tau}\!\equiv\!r_{\tau}(a_0)$, being functions of the mentioned strong
QCD couplant $a_0$. There is, however, an additional freedom
of choosing the value of the constant ${\tilde c}$ in
Eq.~(\ref{rtRjw}).\footnote{
Variation of ${\tilde c}$ in Eq.~(\ref{rtRjw}) corresponds to
changing somewhat the numerical procedure used.}
Since the available series of ${\overline R}_j(b(w_k);{\tilde c})$
is truncated, the results of Eq.~(\ref{rtRjw}) will have some
unphysical dependence on the value of ${\tilde c}$.
In one of our approaches, we will choose the latter value 
by the principle of minimal
sensitivity (PMS), i.e., by the condition
\begin{equation}
 \frac{\partial r_{\tau}(a_0;{\tilde c})}{\partial {\tilde c}} = 0 
\ ,
\label{PMS}
\end{equation}
when using for ${\overline R}_j(b(w_k))$ their
${\rm N}^3 {\rm LO}$ TPS forms.

We can, however, proceed in a different way.
The ${\rm N}^3 {\rm LO}$ TPS of ${\overline R}_j(b(w_k);{\tilde c})$
can be further resummed using 
Pad\'e approximants (PA's) \cite{Padebook}.\footnote{
PA $[n/m]_R(w)$ to a function $R(w)$ is a ratio of
polynomials in $w$ of degree $n$ (numerator) and $m$ (denominator).
The power expansion of $[n/m]_R(w)$ must reproduce
the terms of the power expansion of $R(w)$ up to, and including,
the term $\sim\!w^{n+m}$. PA $[n/m]_R(w)$ can be determined 
if we know the TPS of $R(w)$ up to, and including, the term
$\sim\!w^{n+m}$. PA's $[n/n]$ are called diagonal.}
The authors of Ref.~\cite{Jentschura:2001fm} presented compelling arguments 
that combining the conformal transformations with the PA--type
of resummations in general leads to significantly improved results,
at least when a sufficient number of terms in the power
expansion are known.
Especially the diagonal or almost diagonal PA's 
\cite{Padebook,Jentschura:2001fm}, 
in our case $[2/1]_{{\overline R}_j}(w_k)$ and 
$[1/2]_{{\overline R}_j}(w_k)$,
may represent an efficient way of extending the applicability
of expressions for ${\overline R}_j$ into the region
sufficiently far away from the origin (quasianalytic
continuation). 

However, there is a possibility that PA's
do not lead to an improvement. This is sometimes the case when
the TPS in question is known to a relatively low order, e.g.
to the ${\rm N}^3 {\rm LO}$. Since our available TPS's of
${\overline R}_j(w_k;{\tilde c})$ ($k\!=\!3,4$) are known to the
${\rm N}^3 {\rm LO}$ (provided a specific value of $d_3^{(0)}$ is
taken), we have to find criteria for keeping or rejecting
the resulting PA's. The PA $[2/1]_{{\overline R}_j}(w_k;{\tilde c})$
predicts one real singularity of ${\overline R}_j(b(w_k);{\tilde c})$,
and $[1/2]_{{\overline R}_j}(w_k;{\tilde c})$ two singularities which can
be either real or mutually complex conjugate. Physically,
${\overline R}_j(b(w_k);{\tilde c})$ has the strongest singularities 
(UV and IR renormalon singularities)
at $b(w_k)\!=\!-1, -2, \ldots$ and $b(w_k)\!=\!3,4, \ldots$.
This means that in the case of conformal transformations
(\ref{ct34}), the singularities of PA's should preferably
be at $w_k$ values corresponding to $b\!=\!-1$ or $3$:
$w_3^{\rm pole}\!=\!-1$ or $+1$;
$w_4^{\rm pole}\!=\!-1$ or $+0.6$.
We will include in our analysis the predictions with
those PA's $[2/1]_{{\overline R}_j}(w_k)$ whose
poles satisfy the latter conditions.
For that, we will use our freedom to adjust the
value of the constant ${\tilde c}$ in Eq.~(\ref{rtRjw}).
On the other hand, PA's $[1/2]_{{\overline R}_j}(w_k;{\tilde c})$, 
which have two poles, only rarely satisfy
approximately the aforementioned conditions simultaneously.

Furthermore, in practical calculations, we prefer to
use for the $\beta_{\overline {\rm MS}}$ (\ref{betaMS})
in the Stevenson's equation (\ref{Steve}) the PA
$[2/3]_{\beta {\overline {\rm MS}}}(x)$, particularly because
of its reasonable singularity structure ($x_{\rm pole} \approx 0.311$,
corresponding to $\alpha_s^{\rm pole} \approx 0.98$). The latter signals
the breakdown of perturbative QCD (pQCD), and this choice 
of $\beta_{\overline {\rm MS}}(x)$ has been used previously
by some of us in Refs.~\cite{Cvetic:2001sn,Cvetic:2001mz}.
Since the effective energy in the discussed 
QCD observable $r_{\tau}$ is relatively low $E\!\sim\!m_{\tau}\!<\!2$ GeV,
we can expect that this choice is not entirely irrelevant numerically.
Later we will comment on how much the results change when
employing the (${\rm N}^3 {\rm LO}$) TPS for
$\beta_{\overline {\rm MS}}(x)$, instead. 

\section{Predictions for \lowercase{$\alpha_s$}; theoretical
uncertainties estimate}
\label{sec:predictions}

As discussed in the previous section, 
for a given choice of 
$a_0\!\equiv\!\alpha_s(m^2_{\tau};{\overline {\rm MS}})/\pi$
and of constant ${\tilde c}$,
we calculate the observable $r_{\tau}\!\equiv\!r_{\tau}(a_0;{\tilde c})$ 
(\ref{rtRjw}), using for ${\overline R}_j(b(w_k);{\tilde c})$ 
one of the variants (\ref{barR1})--(\ref{barR3}) ($j\!=\!1,2,3$),
and one of the two conformal transformations
(\ref{ct34inv}) ($k\!=\!3,4$), and for the resummation
of ${\overline R}_j(b(w_k))$ either the ${\rm N}^3 {\rm LO}$ TPS
or $[2/1]$ PA. We thus use, at given ${\tilde c}$,
altogether $3 \times 2 \times 2 = 12$
ways of calculating $r_{\tau}$ as function of
$a_0\!\equiv\!{\alpha}_s(m^2_{\tau},{\overline {\rm MS}})/\pi$.
The value ${\tilde c}$ is adjusted as described at the end
of the previous section: (a) according to the PMS (\ref{PMS})
when using ${\rm N}^3 {\rm LO}$ TPS for ${\overline R}_j(b(w_k))$;
(b) according to the pole requirements 
$w_3^{\rm pole}\!=\!-1$ or $+1$
($w_4^{\rm pole}\!=\!-1$ or $+0.6$) when using PA
$[2/1]_{{\overline R}_j}(w_k)$ for ${\overline R}_j(b(w_k))$.
We repeat the analysis for three choices of the
${\rm N}^3 {\rm LO}$ coefficient $d_3^{(0)}$ (\ref{d3est})
of the Adler function: $d_3^{(0)}\!=\!25$, $15$, $35$
In the central case of $d_3^{(0)}\!=\!25$, and for $r_{\tau}$
close to the central measured value $r_{\tau}\!=\!0.1960$
(\ref{rtauexp}), we display the relevant numerical results
in Table \ref{tabl1}. 
Displayed are the predictions of various
approximants for the input values 
$\alpha_s(m^2_{\tau}; {\overline {\rm MS}}) = 0.325$ and $0.326$.
The first 14 entries are predictions of Eq.~(\ref{rtRjw})
when using PA's $[2/1]_{{\overline R}_j}(w_k)$
($j\!=\!1,2,3$; $k\!=\!3,4$), where ${\tilde c}$ was
adjusted so that these PA's yield the aforementioned
pole values of the leading UV or subleading IR renormalon
pole. The next six entries are predictions when
using ${\rm N}^3 {\rm LO}$ TPS for ${\overline R}_j(b(w_k))$
in Eq.~(\ref{rtRjw}) with ${\tilde c}$ adjusted so that the
PMS principle (\ref{PMS}) is satisfied (local maximum).
The last entry contains the arithmetic average ${\bar r}_{\tau}$
of all 20 predictions, for the aforementioned two values
of $\alpha_s(m^2_{\tau}; {\overline {\rm MS}})$.
From that entry, we deduce that
the central measured value $r_{\tau}\!=\!0.1960$
(\ref{rtauexp}) is achieved by this arithmetic
average ${\bar r}_{\tau}$ at 
$\alpha_s(m^2_{\tau}; {\overline {\rm MS}}) = 0.3254$.
The uncertainty of the prediction due to resummation method
(``truncation'' error) can be estimated
by comparing the aforementioned prediction with
the prediction which differs the most from it,
i.e., with the prediction using
the PA $[2/1]_{{\overline R}_j}(w_k)$ with $j\!=\!3$,
$k\!=\!4$ and ${\tilde c}\!=\!1.34$ (see the 14th
entry of Table \ref{tabl1}). This prediction differs from
the aforementioned one by 
$|\delta \alpha_s|_{\rm tr.} \approx 0.0024$. 

Repeating the very same calculations (at the same
values of ${\tilde c}$) for the correspondingly higher
and lower input values of $\alpha_s(m^2_{\tau}; {\overline {\rm MS}})$,
we obtain the corresponding predictions of
$\alpha_s(m^2_{\tau}; {\overline {\rm MS}})$ for the
upper and lower bounds of the measured values of $r_{\tau}$
(\ref{rtauexp}) by demanding that the arithmetic
average ${\bar r}_{\tau}$ be equal to those upper and lower
bounds. This procedure then results in the prediction
\begin{eqnarray}
\alpha_s^{\overline {\rm MS}}(m^2_{\tau}) & = &
0.3254 \pm 0.0060_{\rm exp.} \pm 0.0060_{\rm EW} \pm 0.0052_{\rm CKM}
\pm 0.0024_{\rm tr.}
\quad (d_3^{(0)}\!=\!25) \ .
\label{al0d325}
\end{eqnarray}
It is gratifying that in the case of
$d_3^{(0)}\!=\!25$ so many PA cases give physically acceptable
pole structure, and that the predictions for
$\alpha_s^{\overline {\rm MS}}(m^2_{\tau})$ of the
aforementioned 20 different approaches
differ from each other only little -- they differ by at most
0.0024 from the prediction of their total average.
If confining ourselves to just one conformal transformation,
the central value changes by only $\pm0.0001$ (0.3255 for
$k\!=\!3$; 0.3253 for $k\!=\!4$). If using only the
six entries from the approach with the PMS, the central value is
$0.3259$. If using only the entries with 
$w_k^{\rm pole} \approx -1$, the central value is 0.3258.
If using only the entries with $w_3^{\rm pole} \approx +1$
and $w_4^{\rm pole} \approx +0.6$, the central value
changes to 0.3244.

Further, for most of the approximants of the
first eight entries of Table \ref{tabl1} (except those
with $j\!=\!1$), when using the same values of
${\tilde c}$ but using PA's $[1/2]$ instead of
$[2/1]$ for ${\overline R}_j(w_k)$, 
the predictions $r_{\tau}$ differ from
those of $[2/1]$ by no more than 0.0005, and
the predictions of $\alpha_s(m^2_{\tau})$ also
by no more than 0.0005. In these cases of good agreement,
the poles $w_k^{\rm pole}$ of the PA's $[1/2]$ do 
not lie deep inside the $b(w_k)$--intervals $[-1,+3]$. 
In most of the cases when the disagreement is larger,
at least one of the poles of the $[1/2]$ falls deep inside
these intervals, often even within the intervals $[-0.5,+2]$.
This offers us an additional evidence that our requirement
that the poles of $[2/1]$ be either at $b(w_k)\!=\!-1$ 
(leading UV renormalon pole) of $b(w_k)\!=\!+3$
(next--to--leading IR renormalon pole) was reasonable.  

In order to obtain the uncertainties of the prediction
due to the uncertainty $d_3^{(0)}\!=\!25 \pm 10$
around the central value $d_3^{(0)}\!=\!25$,
we repeat the procedure for the case of 
$d_3^{(0)}\!=\!15$ and $d_3^{(0)}\!=\!35$.

In the case $d_3^{(0)}\!=\!15$, the situation
is very similar to the afore--discussed
case of $d_3^{(0)}\!=\!25$. We obtain six entries
when we require $w_k^{\rm pole}\!=\!-1$;
three entries each when $w_3^{\rm pole}\!=\!+1$
and $w_4^{\rm pole}\!=\!+0.6$;
six entries when applying the PMS condition (\ref{PMS})
with ${\rm N}^3 {\rm LO}$ TPS for ${\overline R}_j(b(w_k))$
-- see Table \ref{tabl2}, when two choices
${\alpha}_s(m^2_{\tau}; {\overline {\rm MS}})\!=\!0.331$ and
$0.332$ are made.
The arithmetic average ${\bar r}_{\tau}$
of all these 18 entries is then equal to
the central measured value (\ref{rtauexp})
$r_{\tau}\!=\!0.1960$ when 
$\alpha_s(m^2_{\tau}; {\overline {\rm MS}})\!=\!0.3299$,
i.e., the value higher by $0.0045$ than the 
corresponding central prediction of Eq.~(\ref{al0d325}) of the case 
$d_3^{(0)}\!=\!25$.

In the case $d_3^{(0)}\!=\!35$, the numerical situation
is less favorable. None of the approaches with
PA's $[2/1]_{{\overline R}_j}(w_k)$ have acceptable
solutions under the requirement $w_k^{\rm pole}\!=\!-1$.
Such poles occur at ${\tilde c} < -2$, but the
corresponding predictions of $r_{\tau}$ there are
quite unstable under the variation of ${\tilde c}$.
In two cases, acceptable solutions are obtained
with the approach with PA's $[2/1]_{{\overline R}_j}(w_k)$
when we require $w_3^{\rm pole}\!=\!+1$
or $w_4^{\rm pole}\!=\!+0.6$ -- see the first two
entries of Table \ref{tabl3}. 
The approach with the PMS (\ref{PMS}),
when using ${\rm N}^3 {\rm LO}$ TPS for ${\overline R}_j(b(w_k))$,
appears to be more difficult as well; the derivatives
$\partial r_{\tau}/\partial {\tilde c}$ are never zero,
but are negative for all reasonable values of ${\tilde c}$;
nonetheless, these slopes have the smallest negative
values at specific values of ${\tilde c}$ -- see
the corresponding six entries in Table \ref{tabl3}.
Taking the arithmetic average ${\bar r}_{\tau}$
of the eight entries of Table \ref{tabl3}, we infer that
the arithmetic average achieves the central measured value
(\ref{rtauexp}) $r_{\tau}\!=\!0.1960$ for the
value $\alpha_s(m^2_{\tau}; {\overline {\rm MS}})\!=\!0.3194$,
which is lower by $0.0060$ than the 
corresponding central prediction of Eq.~(\ref{al0d325}) of the case 
$d_3^{(0)}\!=\!25$.

{}From the above considerations we infer that the
uncertainty of the prediction of $\alpha_s(m^2_{\tau}; {\overline {\rm MS}})$
due to the uncertainty (\ref{d3est}) of $d_3^{(0)}$
is $\pm0.0060$. 

There is yet another theoretical uncertainty involved in the
prediction (\ref{al0d325}), connected with the choice of the
renormalization scheme (RSch). As shown in the Appendix,
when taking a RSch different from the modified `t Hooft
(mtH) RSch $c_k\!=\!c_1^k$ ($k\!=\!2,3,\ldots$), 
the integral transformations involved
become more complicated. Then we end up with, instead of the
simpler formula  (\ref{rtRjw}), the more complicated
one (\ref{rtaugen}). We note that the leading RSch parameter
$c_2$ has the values $c_2 \approx 3.16; 4.47; 6.58; 5.24$
in the mtH, ${\overline {\rm MS}}$, principle of minimal
sensitivity (PMS) \cite{PMS}, 
and the effective charge (ECH) \cite{ECH} RSch's
(where we take $n_f\!=\!3$; and the PMS and ECH RSch's refer
to the ${\rm N}^3 {\rm LO}$ TPS of the Adler function). 
Therefore, we estimate as a characteristic
deviation from $c_2\!=\!c_1^2 \approx 3.16$ the value
$c_2 \approx c_1^2 + 3.4$. We then perform the analysis
in the RSch with the latter value of $c_2$ (all the other
$c_k$'s unchanged), using formula (\ref{rtaugen}). We use
the RScl parameter value $\xi^2\!\equiv\!\mu^2/Q^2 = 1$
[$Q^2\!=\!m^2_{\tau} \exp( {\rm i} y)$], for $d_3^{(0)}$ 
we use the central value $d_3^{(0)}\!=\!25$, and
for the $\beta$--functions we use again the $[2/3]$ PA form.
The results of the analysis are written in Table
\ref{tabl4}. The variation of the parameter
${\tilde c}$ allowed us to obtain the desired locations
of the poles only in a few cases. The arithmetic average
values ${\overline r}_{\tau}$ of Table \ref{tabl4}
reach the central value prediction
${\overline r}_{\tau}\!=\!0.1960$ when the the coupling is 
$\alpha_s(m^2_{\tau}; {\overline {\rm MS}}) = 0.3285$,
which is by $0.0031$ higher than the central
value in Eq.~(\ref{al0d325}). Concerning the change of
the next--to--leading RSch parameter $c_3$, we note that
$c_3 \approx 5.6; 21.0; 36.8; 16.1$ for the
mtH, ${\overline {\rm MS}}$, PMS and ECH RSch's.
Therefore, we choose a characterictic deviation
of $c_3$ from the mtH value $c_3 = c_1^3 \approx 5.6$
to be $c_3 = c_1^3\!+31$. Completely analogous
analysis as in the case of the $c_2$--deviation
leads to the results of Table \ref{tabl5}.
The arithmetic average values 
${\overline r}_{\tau}$ in Table \ref{tabl5}
reach the central value ${\overline r}_{\tau}\!=\!0.1960$
for the coupling 
$\alpha_s(m^2_{\tau}; {\overline {\rm MS}}) = 0.3274$,
which is by $0.0020$ higher than the central
value in Eq.~(\ref{al0d325}). Therefore, adding the
deviations $0.0031$ and $0.0020$ in quadrature
gives the estimated uncertainty due to the changes
of the RSch to be $\pm 0.0037$.

We can thus add to the prediction
(\ref{al0d325}) the discussed uncertainties due to the
variation of $d_3^{(0)}$ and of the RSch, resulting in
our final prediction
\begin{eqnarray}
\alpha_s^{\overline {\rm MS}}(m^2_{\tau}) & = &
0.3254 \pm 0.0060_{\rm exp.} \pm 0.0060_{\rm EW} \pm 0.0052_{\rm CKM}
\nonumber\\
&& \pm 0.0060_{\delta d_3} \pm 0.0037_{\rm RSch} \pm 0.0024_{\rm tr.} \ ,
\label{res1a}
\\
& = & 0.3254 \pm 0.0060_{\rm exp.} \pm 0.0079_{\rm EW+CKM} \pm 
0.0074_{\rm th.} \ .
\label{res1b}
\end{eqnarray}
In the last line, we added the corresponding uncertainties in
quadrature; the combined uncertainty due to the uncertainty
of $d_3^{(0)}$ (\ref{d3est}), the resummation (``truncation'')
uncertainty, and the RSch uncertainty 
we call the theoretical (th.\/) uncertainty. 
This combined uncertainty
is comparable with the two other uncertainties in Eq.~(\ref{res1b}).
If we use for the ${\overline {\rm MS}}$
$\beta$--function in Eq.~(\ref{Steve}) the ${\rm N}^3 {\rm LO}$ TPS
form instead of the $[2/3]_{\beta {\overline {\rm MS}}}(x)$
PA form, we obtain, in a way completely analogous to that
described in Table \ref{tabl1}, the central value prediction
$\alpha_s(m^2_{\tau}; {\overline {\rm MS}})\!=\!0.3245$.
This is by $0.0009$ lower than the central value
prediction in Eqs.~(\ref{res1a})--(\ref{res1b}), indicating that
those nonperturbative effects which originate from the
behavior of the $\beta$--function are not strong. 
The fact that we used in the Borel integral a
finite $b(w)_{\rm max}$ ($\!=\!4$) does not influence
the results. Namely, if we increase this quantity
to $b(w)_{\rm max}\!=\!5$ [corresponding for $w_3$
to $\phi_3\!=\!0.64350$ in Eq.~(\ref{rtRjw}),
and for $w_4$ to $\phi_4\!=\!0.40272$],
the predictions for $\alpha_s(m^2_{\tau})$
change by $\sim\!10^{-6}$, i.e., insignificantly.

We then RGE--evolved the result (\ref{res1a})--(\ref{res1b})
from the scale $\mu=m_{\tau} \approx 1.777$ GeV to the scale
$M_{\text{z}}\!=\!91.19$ GeV. We used again the aforementioned
$[2/3]_{\beta {\overline {\rm MS}}}(x)$ PA form of the
$\beta_{\overline {\rm MS}}$--function, which is
based on the known
four--loop ${\rm N}^3 {\rm LO}$ TPS form of $\beta_{\overline {\rm MS}}$
\cite{vanRitbergen:1997va}. Therefore, we employed the corresponding
three--loop matching conditions \cite{Chetyrkin:1997sg}
for the flavor thresholds. The matching was performed at
$\mu(N_f)\!=\!\kappa m_q(N_f)$ with the choice $\kappa\!=\!2$,
where $\mu(N_f)$ is the scale
above which $N_f$ flavors are assumed active, and
$m_q(N_f)$ is the running quark mass $m_q(m_q)$
of the $N_f$'th flavor. We further assumed
$m_c(m_c)\!=\!1.25$ GeV and $m_b(m_b)\!=\!4.25$ GeV
\cite{Groom:2000in}.
We thus obtain from Eqs.~(\ref{res1a})--(\ref{res1b})
\begin{eqnarray}
\alpha_s^{\overline {\rm MS}}(M^2_{\text{z}}) &=& 
0.1192 \pm 0.0007_{\rm exp.}
\pm 0.0010_{\rm EW+CKM} \pm 0.0009_{\rm th.} \pm 0.0003_{\rm evol.}
\ , 
\label{res2a}
\\
& = & 0.1192 \pm 0.0015 \ .
\label{res2b}
\end{eqnarray}
In Eq.~(\ref{res2b}), we added all the uncertainties 
in quadrature.
In Eq.~(\ref{res2a}), we included the uncertainties due to
the RGE evolution, which come primarily from varying
$\kappa$ from $1.5$ to $3$, and from varying the quark masses
$m_c(m_c)\!=\!1.25 \pm 0.10$ GeV and $m_b(m_b)\!=\!4.25 \pm 0.15$ GeV
(see Ref.~\cite{Cvetic:2001sn} for more details).

If we repeat the calculation of $\alpha_s(m^2_{\tau})$
and $\alpha_s(M^2_{\rm z})$ with the ${\rm N}^3 {\rm LO}$
TPS $\beta_{\overline {\rm MS}}$ function [in Eq.~(\ref{Steve})
and in the RGE evolution from $m^2_{\tau}$ to $M^2_{\rm z}$], 
instead of the used PA $[2/3]_{\beta {\overline {\rm MS}}}(x)$,
the central value prediction 
$\alpha_s(M^2_{\rm z}; {\overline {\rm MS}})\!=\!0.1192$ 
remains unchanged up to the displayed digits. This is so because
this change of $\beta_{\overline {\rm MS}}$ causes
the central value prediction of 
$\alpha_s(m^2_{\tau}; {\overline {\rm MS}})$
to be by $0.0009$ lower [as already mentioned after 
Eqs.~(\ref{res1a})--(\ref{res1b})], but then the
RGE evolution to $\mu^2\!=\!M^2_{\rm z}$ with the
changed $\beta_{\overline {\rm MS}}$ pushes the
result up, approximately neutralizing the former effect. 

In the analysis leading to the results (\ref{res1a})--(\ref{res2b})
we assumed that the power--suppressed terms, apart from those
from the quark masses, do not contribute to the 
considered observable $R_{\tau}$, as already emphasized
in Sec.~\ref{sec:basic}. As mentioned in that section,
the inclusive $({\rm V}+{\rm A})$ fit of the ALEPH 
Collaboration \cite{Barate:1998uf}, within their
framework, predicted the contributions of the (massless) 
power--suppressed terms to the
canonical observable $r_{\tau}$ to be consistent with
zero: $\delta r_{\tau, {\rm PS}} = 0.000 \pm 0.004$.
If we assumed that the latter estimates were valid also
in our framework,\footnote{
See the discussion in Sec.~\ref{sec:basic} about the
differences between our and ALEPH's framework.}
this $\delta r_{\tau, {\rm PS}}$ 
would have to be subtracted from the
values given on the right-hand side of 
(\ref{rtauexp}), resulting in an additional,
``PS''--uncertainty, term $\pm 0.0040$ for the
$r_{\tau}$. This would in turn give
an additional approximate uncertainty
$\pm 0.0041_{\rm PS}$ in the result
(\ref{res1a})-(\ref{res1b}) for
$\alpha_s(m^2_{\tau})$, and
$\pm 0.0005_{\rm PS}$ in the result
(\ref{res2a}) for $\alpha_s(M^2_{\rm z})$.
The combined uncertainty $\pm0.0015$ for $\alpha_s(M^2_{\rm z})$
in (\ref{res2b}) would increase to $\pm0.0016$. All the central
value predictions in (\ref{res1a})--(\ref{res2b})
would remain unchanged.

There are at least two indications that the above results 
(\ref{res1a})--(\ref{res1b}) and (\ref{res2a})--(\ref{res2b})
are not wrong. In the approach of Ref.~\cite{Cvetic:2001sn},
which involved ordinary  Borel transform of $D(Q^2)$
and where the RScl was fixed according to the (local) PMS
and we used the ${\overline {\rm MS}}$ RSch, the
resulting predictions were very similar:
$\alpha_s(m^2_{\tau}) = 0.3267 \pm 0.0062_{\rm exp.} \pm 0.0082_{\rm EW + CKM} 
\pm 0.0073_{\rm meth.}$
and
$\alpha_s(M^2_{\text{z}}) = 0.1193 \pm 0.0007_{\rm exp.}
\pm 0.0010_{\rm EW+CKM} \pm 0.0009_{\rm meth.} \pm 0.0003_{\rm evol.}$
$= 0.1193 \pm 0.0015$.\footnote{
The method (meth.\/) uncertainty
in \cite{Cvetic:2001sn} is the combination of uncertainties
from $\delta d_3^{(0)}$, the truncation (resummation),
and the RScl and RSch ambiguities; the method uncertainty
there thus corresponds to our theoretical (th.\/) uncertainty
in Eqs.~(\ref{res1b}) and (\ref{res2a}).}
Especially the latter values are in virtual agreement
with Eq.~(\ref{res2b}). 

Yet another indication that the results presented here
are correct comes from repeating the entire resummations,
but this time without employing the conformal transformations 
$b\!=\!b(w)$. The $\beta$--functions were again taken in
the $[2/3]$ PA form.
We carried out the resummation with the Borel integration 
in Eq.~(\ref{rtRj}) again
up to $b_{\max}\!=\!4$. For simplicity, we fixed this
time the ${\tilde c}$ value to ${\tilde c}\!=\!0$.
In each of the cases 
$d_3^{(0)}\!=\!25$, $15$, $35$, we excluded from
the analysis the approaches with those PA's which give
physically unacceptable pole structure, i.e., which
have poles well inside the $b$-interval $[-1,3]$.
Those of the PA's $[2/1]_{{\overline R}_j}(b)$
and $[1/2]_{{\overline R}_j}(b)$ which were not excluded
do not contain poles in the $b$--interval $[-0.8,3]$.

In the case $d_3^{(0)}\!=\!25$, the nonexcluded
approaches were those with $[2/1]_{{\overline R}_j}(b)$ for
$j\!=\!1,2,3$;
and $[1/2]_{{\overline R}_j}(b)$ for $j\!=\!1,2$.
Gratifyingly, the $\alpha_s$--predictions of these 
five approaches differ from each other only little: 
$|\triangle \alpha_s(m^2_{\tau})| < 0.0004$. We took the
arithmetic average ${\bar r}_{\tau}$
of all five $r_{\tau}$--predictions,
and obtained $\alpha_s(m^2_{\tau}) = 0.3257 \pm 0.0061_{\rm exp.}
\pm 0.0061_{\rm EW} \pm 0.0053_{\rm CKM}$ (for $d_3^{(0)}\!=\!25$).

In the case $d_3^{(0)}\!=\!35$, the nonexcluded PA approaches
were those with $[2/1]_{{\overline R}_j}(b)$ for $j\!=\!1,2$.
In addition, we did not exclude the approaches with
the ${\rm N}^3 {\rm LO}$ TPS of ${\overline R}_j(b)$
for $j\!=\!2,3$, because their predictions are close
to the aforementioned PA approaches. We then took the
arithmetic average ${\bar r}_{\tau}$ of all four $r_{\tau}$ predictions
and obtained for the central value
$\alpha_s(m^2_{\tau}) = 0.3191$ [corresponding to the
central value $r_{\tau}\!=\!0.1960$ in Eq.~(\ref{rtauexp})],
which is by $0.0066$ lower than the aforementioned
central value $0.3257$ of the $d_3^{(0)}\!=\!25$ case.

In the case $d_3^{(0)}\!=\!15$, the nonexcluded
approaches were those with $[2/1]_{{\overline R}_1}(b)$ 
and $[1/2]_{{\overline R}_1}(b)$. Using the arithmetic
average ${\bar r}_{\tau}$ of these two $r_{\tau}$ predictions, we obtain
the central value $\alpha_s(m^2_{\tau}) = 0.3293$,
which is by only $0.0036$ higher than in the
$d_3^{(0)}\!=\!25$ case.

This leads us to the following predictions of our
method, when no conformal transformation is used
\begin{eqnarray}
\alpha_s^{\overline {\rm MS}}(m^2_{\tau}) & = &
0.3257 \pm 0.0061_{\rm exp.} \pm 0.0081_{\rm EW + CKM} \pm
0.0066_{\delta d_3} \ ,
\label{res1anct}
\\
\alpha_s^{\overline {\rm MS}}(M^2_{\text{z}}) &=& 
0.1192 \pm 0.0007_{\rm exp.}
\pm 0.0010_{\rm EW+CKM} \pm 0.0008_{\rm \delta d_3} \pm 0.0003_{\rm evol.}
\ ,
\label{res2anct}
\end{eqnarray}
which is in almost complete agreement with the
predictions (\ref{res1a}) and (\ref{res2a}), obtained
by the employment of the two conformal transformations
(\ref{ct34}). In Eqs.~(\ref{res1anct}) and (\ref{res2anct}),
we did not include the uncertainties due to the
resummation (truncation) and due to the RSch ambiguities,
because we regard these two predictions only as an
additional cross--check of our main predictions and 
uncertainty estimates (\ref{res1a})--(\ref{res2b}). 

\section{Summary}
\label{sec:summary}

We calculated the hadronic tau decay width $r_{\tau}$
by employing the contour integral form for this
quantity in the complex momentum plane (\ref{rtmud3=0})
and the modified Borel transform for the associated 
perturbative massless Adler function. 
By choosing a special renormalization scheme 
(modified `t Hooft scheme: $c_k\!=\!c_1^k$, $k\!=\!2,3,\ldots$),
the integrand of the modified Borel transform is
renormalization scale invariant. In our approach,
we explicitly account for the structure of the
leading infrared (IR) renormalon of the Adler function
via the corresponding ansatz. Further, to accelerate
the convergence, i.e., to minimize the resummation
(truncation) uncertainties, we employ two different
conformal transformations which ``map away'' all the renormalon
singularities, except the leading and subleading IR
renormalons, onto the unit circle. The correct location of the
leading ultraviolet (UV) renormalon or of the subleading
IR renormalon is enforced by employing Pad\'e approximants
for the truncated perturbation series of the functions
associated with the modified Borel transform.
The Borel integration, in this appraoch, turns out to have 
suppressed renormalon ambiguity for $r_{\tau}$ at the 
leading IR renormalon singularity,
and the ambiguity due to the subleading renormalons
is strongly suppressed by the
exponent in the Borel integral. We neglect 
in the observable $r_{\tau}$ all the possible
power correction terms (except the dimension $d\!=4\!$ quark
mass terms), because the results of the ALEPH analysis
\cite{Barate:1998uf} suggest that such terms
are consistent with zero or negligibly small
even in our resummation framework.

Our analysis predicts the values of 
$\alpha_s^{\overline {\rm MS}}(m^2_{\tau})$
and $\alpha_s^{\overline {\rm MS}}(M^2_{\rm z})$ given in
Eqs.~(\ref{res1a})--(\ref{res2b}). These predictions agree
well with the results obtained in our previous
analysis \cite{Cvetic:2001sn} of $r_{\tau}$ where we employed 
the ordinary Borel transforms. The latter transforms have
significantly different behavior, expansions and the
strengths of the renormalon singularities, and their
integrands in the Borel integral are, in contrast to the present 
approach, renormalization scale dependent. Therefore, our
present predictions represent a powerful reconfirmation of the
predictions of \cite{Cvetic:2001sn}. We consider this to be
important, because analyses of $r_{\tau}$ which do not
involve Borel transforms and do not account for the
leading renormalon structure of the associated Adler
functions \cite{Barate:1998uf,Groote:1998cn,Korner:2001xk,Steele:1998ma,Geshkenbein:2001mn}
give predictions for $\alpha_s(m^2_{\tau})$ and $\alpha_s(M^2_{\rm z})$
which significantly differ from our predictions and
significantly differ among themselves, 
as already emphasized in \cite{Cvetic:2001sn}.
On the other hand, if accounting for the renormalon structure
via a large--$\beta_0$ resummation of the ordinary Borel transform
and employing an ECH--related resummation of $r_{\tau}$,
as performed by the authors of 
Ref.~\cite{Maxwell:2001uv}, their predicted
values $\alpha_s^{\overline {\rm MS}}(M^2_{\rm z}) = 0.120 \pm 0.002$
come significantly closer to our prediction 
(\ref{res2a})--(\ref{res2b}). Further, our prediction 
$\alpha_s^{\overline {\rm MS}}(M^2_{\rm z}) = 0.1192 \pm 0.0015$
is completely compatible with the world average
$0.1184 \pm 0.0031$ as given in Ref.~\cite{Bethke:2000ai},
but somewhat less compatible with the world average
$0.1173 \pm 0.0020$ as given in Ref.~\cite{Hinchliffe:2000yq}.

\acknowledgments

The work of G.C., C.D. and I.S. was supported by FONDECYT (Chile),
Grant No. 1010094 (G.C.) and 8000017 (C.D. and I.S.).
T.L. was supported  by the BK21 Core Project.

\begin{appendix}

\section[]{Modified Borel transforms in the general renormalization scheme}
\setcounter{equation}{0}

\begin{figure}[ht]
 \centering\epsfig{file=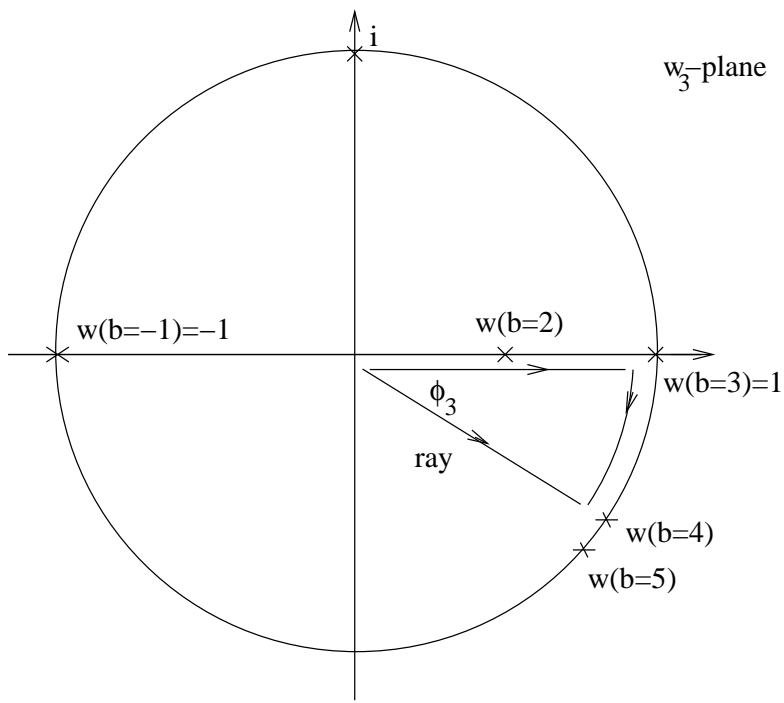}
\vspace{0.3cm}
\caption{\footnotesize
Integration in the $w_3$--plane
along the ray $w_3\!=\!x \exp(-{\rm i} \phi_3)$
($0\!<\!x\!<\!1$, $\phi_3\!=\!0.50536$)
gives the same result as the integration parallel to the
positive real axis ($0 < w < 1$) and arc
$w = \exp(-{\rm i} {\phi}^{\prime})$ ($0 < {\phi}^{\prime} < \phi_3$).
}
\label{fig1}
\end{figure}

\begin{figure}[ht]
 \centering\epsfig{file=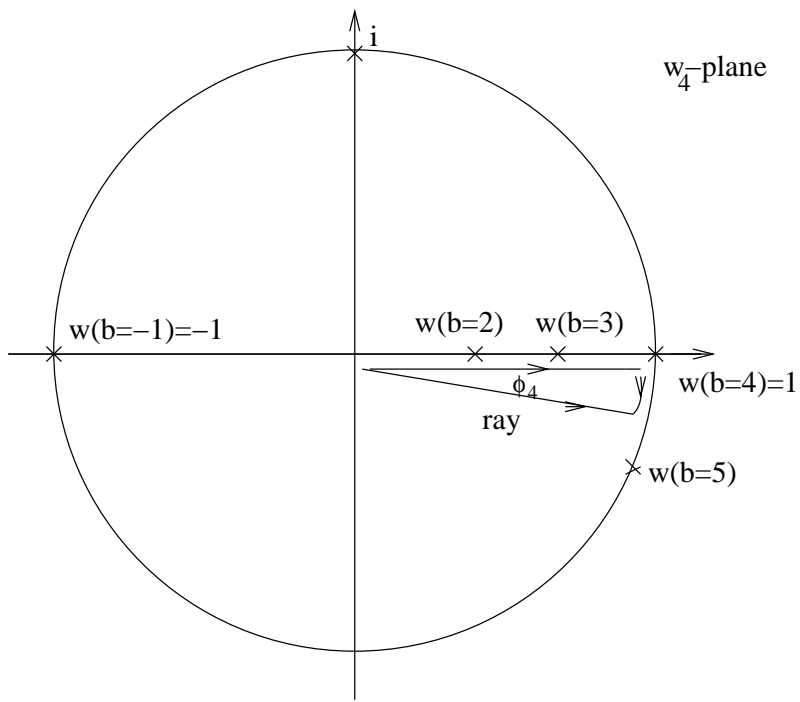}
\vspace{0.3cm}
\caption{\footnotesize
Integration in the $w_4$--plane
along the ray $w_4\!=\!x \exp(-{\rm i} \phi_4)$
($0\!<\!x\!<\!1$, $\phi_4\!=\!0.1$).
}
\label{fig2}
\end{figure}

For a function $f(y)$ with the (asymptotically
divergent) expansion around $y\!=\!0$
\begin{equation}
f(y) = 1 + \sum_{n=1}^{\infty} f_n \;y^n \ ,
\label{fexp}
\end{equation}
the modified Borel transform ${\cal F}_f(\zeta)$ 
was introduced by the authors of \cite{modBT}
via the following expansion:
\begin{equation}
{\cal F}_f(z) = 1 + \sum_{n=1}^{\infty} f_n
\frac{1}{(n\!+\!c_1 z) (n\!-\!1\!+\!c_1 z) \cdots (1\!+\!c_1 z)} z^n \ ,
\label{calFexp}
\end{equation}
where $c_1$ is the coefficient at the two--loop term of the
$\beta$--function (\ref{aRGE}).
The authors of \cite{modBT} further showed that the following
integral transformation connects $f(y)$ and ${\cal F}_f(z)$:
\begin{equation}
f(y) =  \frac{1}{y} (1 - c_1 y) \int_0^{\infty} dz\; 
{\rm e}^{-z/y} \left( \frac{y}{z} \right)^{-c_1 z} 
\frac{1}{\Gamma (1\!+\!c_1 z)} {\cal F}(z) \ .
\label{mBT1}
\end{equation}
Further, they showed that there corresponds to each singularity
$\sim\!(R\!-\!z)^{-\alpha - c_1 R}$ of the ordinary Borel transform
$F_f(z) = 1 + \sum f_n z^n/n!$ a singularity
$\sim\!(R\!-\!z)^{-\alpha}$ of the modified Borel tranform
${\cal F}_f(z)$
\begin{eqnarray}
F_f(z) &=& 1 + \sum_{n=1}^{\infty} \frac{f_n}{n!} \; z^n 
\sim (R - z)^{-\alpha - c_1 R}
\quad \Rightarrow  
\nonumber\\
{\cal F}_{f}(z) &\sim & (R - z)^{-\alpha} \left\{ 1 +
{\cal O} \left[ (R\!-\!z) \ln (1 \!-\!z/R) \right] \right\} \ .
\label{sing}
\end{eqnarray}
On the other hand, the perturbative expansion of the massless Adler 
function $D(Q^2)$ has the canonical form (\ref{Dexpan}), with
$a = a(\mu^2; c_2, c_3, \ldots)$ being the QCD couplant, and the
coefficients $d_k$ having a specific RScl and RSch dependence
$d_k = d_k(\xi^2; c_2, \ldots, c_k)$ ($\xi^2\!\equiv\!\mu^2/Q^2)$
determined by the requirement of the RScl and RSch independence
of $D(Q^2)$ (cf. also \cite{PMS}, first entry)
\begin{eqnarray}
d_1 &=& d_1^{(0)} + \beta_0 \ln \xi^2 \ ,
\label{d1xi2}
\\
d_2 &=& d_2^{(0)} + (d_1^2\!-\!d_1^{(0)2}) + 
c_1 (d_1\!-\!d_1^{(0)}) - (c_2\!-\!c_2^{(0)}) \ ,
\label{d2xi2}
\\
d_3 &=& d_3^{(0)} + 
3 (d_1 d_2\!-\!d_1^{(0)} d_2^{(0)}) - 2 (d_1^3\!-\!d_1^{(0)3})
-(c_1/2) (d_1^2\!-\!d_1^{(0)2}) 
\nonumber\\
&& + (c_2 d_1\!-\!c_2^{(0)}d_1^{(0)}) - (1/2) (c_3 - c_3^{(0)}) \ ,
\label{d3xi2}
\\
d_4 &=& d_4^{(0)} + 4 ( d_1 d_3\!-\!d_1^{(0)} d_3^{(0)} ) -
(c_1/3) (d_3\!-\!d_3^{(0)}) + (5/3)(d_2^2\!-\!d_2^{(0)2})
\nonumber\\
&& - (28/3)(d_1^2 d_2 - d_1^{(0)2} d_2^{(0)}) + 
(2 c_1/3)(d_1 d_2\!-\!d_1^{(0)} d_2^{(0)}) + 
(1/3)(c_2 d_2\!-\!c_2^{(0)} d_2^{(0)})
\nonumber\\
&& + (14/3) (d_1^4\!-\!d_1^{(0)4}) - 
(4/3)(c_2 d_1^2\!-\!c_2^{(0)} d_1^{(0)2}) +
(c_3 d_1\!-\!c_3^{(0)} d_1^{(0)}) - (1/3)(c_4\!-\!c_4^{(0)}) \ .
\label{d4xi2}
\end{eqnarray}
Here, the coefficients $d_k^{(0)}$ are at the RScl $\mu^2\!=\!Q^2$
($\xi^2\!=\!1$) and in the RSch $c_2^{(0)}, c_3^{(0)}, \ldots$.
Comparing expansions (\ref{Dexpan}) and (\ref{fexp}), and inspecting
the integral transformation (\ref{mBT1}), we may identify 
\begin{equation}
a = y; \quad D(Q^2) = \frac{y f(y)}{(1\!-\!c_1 y)} \ ,
\label{corresp}
\end{equation}
which immediately leads to the relations $f_1\!=\!d_1\!-\!c_1$,
$f_k\!=\!d_k\!-\!c_1 d_{k-1}$ ($k\geq 2$). If we use 
$b\!\equiv\!\beta_0 z$ as the Borel variable, this allows us to 
write expansion (\ref{calFexp}) as
\begin{eqnarray}
\lefteqn{
{\cal F}_D(b; \xi^2; c_2, c_3, \ldots) = 
{\Bigg \{} 1 + 
\frac{(d_1\!-\!c_1)}{(1\!+\!c_1 b/\beta_0)} 
\left( \frac{b}{\beta_0} \right)
}
\nonumber\\
&& + \sum_{n=2}^{\infty} \frac{ ( d_n\!-\!c_1 d_{n-1} ) }
{(1\!+\!c_1 b/\beta_0) (2\!+\!c_1 b/\beta_0) \cdots (n\!+\!c_1 b/\beta_0) } 
\left( \frac{b}{\beta_0} \right)^n {\Bigg \}} \ , 
\label{calFDexp}
\end{eqnarray}
and the integral transformation (\ref{mBT1}) as
\begin{eqnarray}
D(Q^2) &=& \frac{1}{\beta_0} \int_0^{\infty} db 
\exp \left[ - \frac{b}{\beta_0 a} \right] 
\left( \frac{ a \beta_0}{b} \right)^{- c_1 b/\beta_0} 
\frac{1}{ \Gamma (1\!+\!c_1 b/\beta_0) }
{\cal F}_D(b) \ ,
\label{mBT2}
\\
&=& 
\frac{1}{\beta_0} \int_0^{\infty} db \exp \left[
- \frac{b}{\beta_0} \left( \frac{1}{a}\!+\!c_1 \ln(c_1 a) \right) \right]
\left( \frac{c_1 b}{\beta_0} \right)^{c_1 b/\beta_0}
\frac{1}{ \Gamma (1\!+\!c_1 b/\beta_0) }
{\cal F}_D(b) \ .
\label{mBT3}
\end{eqnarray}
If the ordinary Borel transform 
${\widetilde D}(b; \xi^2; c_2, \ldots)$ (\ref{SBexp})
has a singularity of the form 
$\sim\!(R\!-\!b/\beta_0)^{-\alpha - c_1 R}$, then
$F_f(z)\!\sim\!(R\!-\!z)^{-\alpha - c_1 R}[1 + 
{\cal O}((R\!-\!z)^1)]$, due to the simple
relation $d F_f/dz = d {\widetilde D}/dz - c_1 {\widetilde D}$.
Therefore, we can write the singularity relation
between the ordinary Borel transform
${\widetilde D}$ and the modified Borel transform
${\cal F}_D\!\equiv\!{\cal F}_f$ in complete analogy with Eq.~(\ref{sing})
\begin{eqnarray}
{\widetilde D}(b; \xi^2; c_2, \ldots) & \sim & 
(\kappa\!-\!b)^{-\alpha - c_1 \kappa/\beta_0}
\quad \Rightarrow 
\nonumber\\
{\cal F}_D(b; \xi^2; c_2, \ldots)  & \sim & 
(\kappa\!-\!b)^{-\alpha} 
\left\{ 1 +
{\cal O} \left[ (\kappa\!-\!b) \ln (1 \!-\!b/\kappa) \right] \right\} \ ,
\label{sing2}
\end{eqnarray}
where $\kappa\!\equiv\!\beta_0 R$.
We may use the subtracted Stevenson equation \cite{PMS}
[cf.~Eq.~(\ref{Steve})] to reexpress (partly)
the expression $(1/a + c_1 \ln (c_1 a))$ in the exponential of
Eq.~(\ref{mBT3}) in terms of $\ln(\mu^2/Q^2)\!\equiv\!\ln(\xi^2)$
and of the invariant $\rho_1(Q^2)$ (\ref{rho1})
\begin{eqnarray}
\frac{1}{a} + c_1 \ln (c_1 a) & = & \rho_1(Q^2) + d_1^{(0)} 
+ \beta_0 \ln \xi^2 -
\int_0^a \left[ \frac{(1\!-\!c_1 x)}{x^2} + 
\frac{\beta_0}{\beta(x)} \right] \ ,
\label{subtrS}
\end{eqnarray}
where $\beta(x)$ is in the RSch considered, i.e., its
expansion around $x\!=\!0$ is: 
$\beta(x;c_2,\ldots)/\beta_0\!=\!-x^2 (1\!+\!c_1 x\!+\!c_2 x^2\!+\!\cdots)$.
Therefore, Eq.~(\ref{mBT3}) can be rewritten
\begin{eqnarray}
D(Q^2) & \equiv & 
D\left( a(\xi^2 Q^2; c_2,\ldots); \xi^2; c_2, \ldots \right) =
\frac{1}{\beta_0} \int_0^{\infty} db 
\exp \left[ - \frac{(\rho_1(Q^2) + \beta_0 \ln \xi^2) b}{\beta_0} \right]
\nonumber\\
&& \times \exp\left[\frac{b}{\beta_0} \int_0^a dx \left(
\frac{(1\!-\!c_1 x)}{x^2} + \frac{\beta_0}{\beta(x;c_2,\ldots)} \right)
\right] 
\nonumber\\
&& \times 
\frac{(c_1 b/\beta_0)^{c_1 b/\beta_0}}{ \Gamma (1\!+\!c_1 b/\beta_0)}
\exp\left(- \frac{d_1^{(0)} b}{\beta_0} \right) 
{\cal F}_D(b; \xi^2; c_2,\ldots) \ .
\label{mBT4}
\end{eqnarray}
In the general RSch $(c_2,c_3,\ldots)$ and at the general RScl
$\xi^2\!\equiv\!\mu^2/Q^2$, the modifed Borel transform function
${\cal F}_D(b; \xi^2; c_2, \ldots)$ is related to the
function ${\cal F}_D(b; 1; c_2^{(0)}, \ldots)$ in the following way:
\begin{eqnarray}
\lefteqn{
{\rm e}^{- b \ln \xi^2} {\cal F}_D(b; \xi^2; c_2, c_3, \ldots) =
{\cal F}_D(b; 1; c_2^{(0)}, c_3^{(0)}, \ldots) - \frac{1}{2}
(c_2\!-\!c_2^{(0)}) \left( \frac{b}{\beta_0} \right)^2 
}
\nonumber\\
&& + \frac{1}{12} \left[ 2 (c_2\!-\!c_1^2) (\beta_0 \ln \xi^2) 
+ (c_2\!-\!c_2^{(0)})(11 c_1\!-\!4 d_1^{(0)}) -
(c_3\!-\!c_3^{(0)}) \right] \left(\frac{b}{\beta_0} \right)^3 
\nonumber\\
&&
+ \frac{1}{72} {\Big [} 3 (c_2\!-\!c_1^2) (\beta_0 \ln \xi^2)
(2 d_1^{(0)}\!-\!\beta_0 \ln \xi^2) +
(22 c_1^3\!+\!3 c_3\!-\!25 c_1 c_2) (\beta_0 \ln \xi^2) 
\nonumber\\
&& 
- (c_2\!-\!c_2^{(0)})(85 c_1^2\!-\!4 c_2\!+\!5 c_2^{(0)}
\!-\!50 c_1 d_1^{(0)}\!+\!9 d_2^{(0)}) +
(c_3\!-\!c_3^{(0)})(13 c_1\!-\!3 d_1^{(0)}) -
(c_4\!-\!c_4^{(0)}) {\Big ]} \left( \frac{b}{\beta_0} \right)^4 
\nonumber\\
&&
+ {\cal {O}}(b^5) \ .
\label{calFDexp2}
\end{eqnarray}
This can be shown, for example, by using transformation
formulas (\ref{d1xi2})--(\ref{d4xi2}) in the expansion
(\ref{calFDexp}). The above formulas (\ref{calFDexp2})
and (\ref{mBT4}) show that the modified `t Hooft (mtH) scheme 
($c_k\!=\!c_1^k$, $k\!=\!2,3,\ldots$) is a remarkable RSch choice
in the discussed Borel transforms
\begin{eqnarray}
{\rm e}^{- b\ln \xi^2} {\cal F}_D(b; \xi^2; c_1^2, c_1^3, \ldots) & = &
{\cal F}_D(b; 1; c_1^2, c_1^3, \ldots) \ ,
\label{calFmtH}
\\
\frac{ (1\!-\!c_1 x) }{x^2} + \frac{\beta_0}{\beta(x;c_1^2,c_1^3,\ldots)}
& = & 0 \ ,
\label{expintmtH}
\end{eqnarray}
and thus Eq.~(\ref{mBT4}) reduces in this case to
\begin{eqnarray}
D(Q^2) & \equiv & D \left( 
a(\xi^2 Q^2; c_1^2, c_1^3 \ldots); \xi^2; c_1^2, c_1^3, \ldots \right) =
\frac{1}{\beta_0} \int_0^{\infty} db 
\exp \left[ - \frac{(\rho_1(Q^2) + \beta_0 \ln \xi^2) b}{\beta_0} \right]
\nonumber\\
&& \times 
\frac{(c_1 b/\beta_0)^{c_1 b/\beta_0}}{ \Gamma (1\!+\!c_1 b/\beta_0)}
\exp\left(\frac{(\beta_0 \ln \xi^2\!-\!d_1^{(0)}) b}{\beta_0} \right) 
{\cal F}_D(b; 1; c_1^2, c_1^3, \ldots) \ .
\label{mBT5}
\end{eqnarray}
This is just the integral transformation (\ref{invBT}), with
expansion (\ref{barDexp}) and the constant 
${\tilde c} = \beta_0 \ln \xi^2$ [note that 
${\tilde d}_1\!=\!d_1^{(0)}\!=\!d_1(\xi^2\!=\!1)$]
\begin{equation}
{\overline D}(b; {\tilde c}) = {\rm e}^{{\tilde c} b/\beta_0}
{\overline D}(b; 0) = 
\frac{(c_1 b/\beta_0)^{c_1 b/\beta_0}}{ \Gamma (1\!+\!c_1 b/\beta_0)}
\exp\left(\frac{({\tilde c}\!-\!d_1^{(0)}) b}{\beta_0} \right) 
{\cal F}_D(b; 1; c_1^2, c_1^3, \ldots) 
{\Big |}_{{\tilde c}= \beta_0 \ln \xi^2} \ .
\label{barDcalF}
\end{equation}
The above expression (\ref{mBT5}) shows the remarkable
property of the mtH RSch: the whole integrand in the 
modified Borel transformation in the mtH RSch is RScl--independent
($\xi^2$--independent). This appears the main reason why
Grunberg \cite{Grunberg:1993hf} called this Borel transformation 
``renormalization scheme invariant.'' Strictly speaking,
we see that this transformation may be called
``renormalization scale (RScl) invariant,'' with the
choice of the mtH renormalization scheme (RSch).
On the other hand, if we abandon the mtH RSch in the
general class of Borel transformations (\ref{mBT4}),
the integrand becomes explicitly RScl--dependent
\begin{eqnarray}
\lefteqn{
\frac{\partial [ {\rm integrand(b; \xi^2; c_2, \ldots)} ] }
{\beta_0 \partial \ln \xi^2}  
} 
\nonumber\\
&\propto&
(c_1 b/\beta_0)^{c_1 b/\beta_0} \left\{
\left( \frac{b}{\beta_0} \right) \left[
1 + \frac{(1\!-\!c_1 a) \beta(a; c_2, \ldots)}{\beta_0 a^2}
\right] + {\cal {O}}(b^2) \right\}
\nonumber\\
& = & (c_1 b/\beta_0)^{c_1 b/\beta_0} \left\{
\left( \frac{b}{\beta_0} \right) \left[
1 - (1\!-\!c_1 a)(1\!+\!c_1 a\!+\!c_2 a^2\!+ \cdots) \right]
+ {\cal {O}}(b^2) \right\} \ .
\label{intRScl}
\end{eqnarray}
In order to obtain the expression for $r_{\tau}$ in terms
of the modified Borel transform ${\cal F}_D(b; \xi^2; c_2, \ldots)$
in the general RSch, the contour integration (\ref{rtmud3=0})
in the complex momentum plane has to be performed
on the massless Adler function expression (\ref{mBT4}).
Further, we can perform in addition the conformal transformation
$b\!=\!b(w)$ of the types (\ref{ct34inv}) and the ray
integration trick in the $w$--plane as explained 
in Figs.~\ref{fig1},\ref{fig2}.
The procedure is analogous to the procedure leading to
formula (\ref{rtRjw}) in the mtH RSch, and we end up with
the following formula in the general RSch:
\begin{eqnarray}
r_{\tau} & = & \frac{1}{2 \pi \beta_0} {\rm Re} {\Bigg \{}
e^{- {\rm i} \phi} \int_0^1 dx \; \frac{d b(w)}{d w} \;
\exp \left[ 
- \frac{ \left({\rho}_1(m^2_{\tau};a_0)\!+\!{\tilde c} \right)
b(w)}{\beta_0} \right] 
\nonumber\\
&& \times 
\frac{\left( c_1 b(w)/\beta_0 \right)^{c_1 b(w)/\beta_0} 
e^{ - c_1 b(w)/\beta_0} }
{ \Gamma \left( 1\!+\!c_1 b(w)/\beta_0 \right) } 
f_j(b(w)) {\overline R}_j(b(w); {\tilde c}; \xi^2; c_2, \ldots)
\nonumber\\
&& \times
\frac{1}{(1\!-\!b(w)/2)}
\int_{-\pi}^{\pi} dy  (1\!+\!e^{{\rm i}y})^3 (1\!-\!e^{{\rm i} y})
e^{-{\rm i} b(w) y}  
\nonumber\\
&& \times
\exp \left[ \frac{b(w)}{\beta_0} 
\int_0^{a(\xi^2 m^2_{\tau} \exp({\rm i}y); c_2, \ldots)}
dx \left( \frac{(1\!-\!c_1 x)}{x^2} + \frac{\beta_0}{\beta(x; c_2, \ldots)}
\right) \right] {\Bigg \}} {\Bigg |}_{w=x e^{{\rm i} \phi}} \ ,
\label{rtaugen}
\end{eqnarray}
where the weight functions $f_j$ are given in Eq.~(\ref{fjs})
and the Borel transform functions ${\overline R}_j$
are defined in analogy with Eqs.~(\ref{barR1})--(\ref{barR3})
\begin{eqnarray}
{\overline R}_1(b; {\tilde c}; \xi^2; c_2, \ldots) & = &
(1\!-\!b/2) \exp[({\tilde c}\!+\!c_1\!-\!d_1^{(0)}) b/\beta_0]
{\rm e}^{- b \ln \xi^2} {\cal F}_D(b; \xi^2; c_2, c_3, \ldots) \ ,
\label{barR1gen}
\\
{\overline R}_2(b; {\tilde c}; \xi^2; c_2, \ldots) & = &
{\overline R}_1(b; {\tilde c}; \xi^2; c_2, \ldots) 
\frac{(1 + c_1 b/\beta_0)}{(2 + c_1 b/\beta_0)} \ ,
\label{barR2gen}
\\
{\overline R}_3(b;{\tilde c}; \xi^2; c_2, \ldots) & = & 
{\overline R}_1(b;{\tilde c}; \xi^2; c_2, \ldots) 
\frac{(1 + c_1 b/\beta_0)(2 + c_1 b/\beta_0)}
{(3 + c_1 b/\beta_0)^2} \ .
\label{barR3gen}
\end{eqnarray}
We introduced an additional freedom factor $\exp(- {\tilde c}\;b/\beta_0)$
in the exponential in Eq.~(\ref{rtaugen}), which is then offset
by the factor $\exp(+ {\tilde c}\;b/\beta_0)$ in the
functions ${\overline R}_j$, in analogy with the
case of the mtH RSch Eq.~(\ref{rtRjw}). The functions
${\overline R}_j(b(w);\xi^2; {\tilde c}; c_2, \ldots)$
can be resummed, either as PA $[2/1]_{{\overline R}_j}(w)$
or as simple ${\rm N}^3 {\rm LO}$ TPS. The function
$\exp(- b \ln \xi^2) {\cal F}_D(b; \xi^2; c_2, c_3, \ldots)$,
which is the source of the RScl--dependence in
${\overline R}_j$'s, has only a weak RScl--dependence,
as can be seen from Eq.~(\ref{calFDexp2})
\begin{eqnarray}
e^{- b \ln \xi^2} {\cal F}_D(b; \xi^2; c_2, c_3, \ldots)
& = & {\cal F}_D(b; 1; c_2, c_3, \ldots) +
\nonumber\\
&& \frac{1}{6} (c_2\!-\!c_1^2) \left( \frac{b}{\beta_0} \right)^3
( \beta_0 \ln \xi^2) + {\cal {O}}(b^4) \ .
\label{calFxi2}
\end{eqnarray}
We further see from Eq.~(\ref{rtaugen}) that in the general RSch
we cannot perform the contour integration over $dy$ analytically,
in contrast to the mtH RSch Eq.~(\ref{rtRjw}).

\end{appendix}

\noindent

\noindent

\begin{table}[ht]
\par
\begin{center}
\begin{tabular}{c c  c  c}
($j$, $k$); approximant & 
$r_{\tau}$ [$\alpha_s(m^2_{\tau})\!=\!0.325; 0.326$] &
${\tilde c}$ &  comments \\
\hline \hline
$(1,3)$; $[2/1]$ & 0.19543; 0.19641 & +0.28 & $w^{\rm pole} \approx -1.$ \\
\hline
$(1,4)$; $[2/1]$ & 0.19524; 0.19621 & +0.17 & $w^{\rm pole} \approx -1.$ \\
\hline
$(2,3)$; $[2/1]$ & 0.19533; 0.19631 & -0.27 & $w^{\rm pole} \approx -1.$ \\
\hline
$(2,4)$; $[2/1]$ & 0.19531; 0.19629 & -0.28 & $w^{\rm pole} \approx -1.$ \\
\hline
$(3,3)$; $[2/1]$ & 0.19570; 0.19668 & -0.15 & $w^{\rm pole} \approx -1.$ \\
\hline
$(3,4)$; $[2/1]$ & 0.19554; 0.19651 & -0.22 & $w^{\rm pole} \approx -1.$ \\
\hline\hline
$(2,3)$; $[2/1]$ & 0.19470; 0.19567 & -0.71 & $w^{\rm pole} \approx -1.$ \\
\hline
$(2,4)$; $[2/1]$ & 0.19466; 0.19563 & -0.89 & $w^{\rm pole} \approx -1.$ \\
\hline\hline
$(1,3)$; $[2/1]$ & 0.19503; 0.19600 & +1.95 & $w^{\rm pole} \approx +1.$ \\
\hline
$(1,4)$; $[2/1]$ & 0.19587; 0.19686 & +2.275 & $w^{\rm pole} \approx +0.6$ \\
\hline
$(2,3)$; $[2/1]$ & 0.19648; 0.19748 & +1.23 & $w^{\rm pole} \approx +1.$ \\
\hline
$(2,4)$; $[2/1]$ & 0.19746; 0.19847 & +1.61 & $w^{\rm pole} \approx +0.6$ \\
\hline
$(3,3)$; $[2/1]$ & 0.19686; 0.19786 & +0.97 & $w^{\rm pole} \approx +1.$ \\
\hline
$(3,4)$; $[2/1]$ & 0.19792; 0.19895 & +1.34 & $w^{\rm pole} \approx +0.6$ \\
\hline\hline
$(1,3)$; TPS & 0.19449; 0.19546 & +0.92 & local max. \\
\hline
$(1,4)$; TPS & 0.19433; 0.19529 & +0.54 & local max. \\
\hline
$(2,3)$; TPS & 0.19545; 0.19643 & +0.47 & local max. \\
\hline
$(2,4)$; TPS & 0.19516; 0.19614 & +0.12 & local max. \\
\hline
$(3,3)$; TPS & 0.19568; 0.19666 & +0.31 & local max. \\
\hline
$(3,4)$; TPS & 0.19535; 0.19633 & -0.01 & local max. \\
\hline\hline
arithm. average ${\bar r}_{\tau}$ & 0.19560; 0.19658 &    & 
\end{tabular}
\end{center}
\caption {\footnotesize The results $r_{\tau}$ of calculations
according to Eq.~(\ref{rtRjw}), using $d_3^{(0)}\!=\!25$, 
employing the conformal transformations
(\ref{ct34inv}) ($k\!=\!3,4$) and the resummations
of ${{\overline R}_j}(w_k)$ (\ref{barR1})--(\ref{barR3}) ($j\!=\!1,2,3$)
as PA $[2/1]$ and as (${\rm N}^3 {\rm LO}$) TPS. The constant
${\tilde c}$ was varied so as to achieve either $w_k^{\rm pole}\!=\!-1$
or $w_3^{\rm pole}\!=\!+1$ ($w_4^{\rm pole}\!=\!+0.6$) when using
PA's $[2/1]$, or stationarity according to the PMS condition
(\ref{PMS}) when using (${\rm N}^3 {\rm LO}$) TPS. The results
$r_{\tau}$ are given for the choices 
$\alpha_s(m^2_{\tau}; {\overline {\rm MS}})\!=\!0.325$ and $0.326$.
The central measured value $r_{\tau}\!=\!0.1960$ (\ref{rtauexp})
is achieved by the arithmetic average ${\bar r}_{\tau}$ at 
$\alpha_s(m^2_{\tau}; {\overline {\rm MS}})\!=\!0.3254$.
}
\label{tabl1}
\end{table}

\begin{table}[ht]
\par
\begin{center}
\begin{tabular}{c c  c  c}
($j$, $k$); approximant & 
$r_{\tau}$ [$\alpha_s(m^2_{\tau})\!=\!0.331; 0.332$] &
${\tilde c}$ &  comments \\
\hline \hline
$(1,3)$; $[2/1]$ & 0.19625; 0.19717 & +1.20 & $w^{\rm pole} \approx -1.$ \\
\hline
$(1,4)$; $[2/1]$ & 0.19600; 0.19692 & +1.10 & $w^{\rm pole} \approx -1.$ \\
\hline
$(2,3)$; $[2/1]$ & 0.19712; 0.19806 & +0.83 & $w^{\rm pole} \approx -1.$ \\
\hline
$(2,4)$; $[2/1]$ & 0.19688; 0.19781 & +0.74 & $w^{\rm pole} \approx -1.$ \\
\hline
$(3,3)$; $[2/1]$ & 0.19776; 0.19871 & +0.55 & $w^{\rm pole} \approx -1.$ \\
\hline
$(3,4)$; $[2/1]$ & 0.19750; 0.19844 & +0.46 & $w^{\rm pole} \approx -1.$ \\
\hline\hline
$(1,3)$; $[2/1]$ & 0.19601; 0.19693 & +2.52 & $w^{\rm pole} \approx +1.$ \\
\hline
$(1,4)$; $[2/1]$ & 0.19687; 0.19781 & +2.83 & $w^{\rm pole} \approx +0.6$ \\
\hline
$(2,3)$; $[2/1]$ & 0.19758; 0.19853 & +1.98 & $w^{\rm pole} \approx +1.$ \\
\hline
$(2,4)$; $[2/1]$ & 0.19856; 0.19953 & +2.30 & $w^{\rm pole} \approx +0.6$ \\
\hline
$(3,3)$; $[2/1]$ & 0.19831; 0.19927 & +1.64 & $w^{\rm pole} \approx +1.$ \\
\hline
$(3,4)$; $[2/1]$ & 0.19942; 0.20040 & +1.97 & $w^{\rm pole} \approx +0.6$ \\
\hline\hline
$(1,3)$; TPS & 0.19543; 0.19634 & +1.6 & local max. \\
\hline
$(1,4)$; TPS & 0.19525; 0.19616 & +1.3 & local max. \\
\hline
$(2,3)$; TPS & 0.19666; 0.19759 & +1.2 & local max. \\
\hline
$(2,4)$; TPS & 0.19640; 0.19733 & +0.9 & local max. \\
\hline
$(3,3)$; TPS & 0.19732; 0.19826 & +0.9 & local max. \\
\hline
$(3,4)$; TPS & 0.19703; 0.19797 & +0.6 & local max. \\
\hline\hline
arithm. average ${\bar r}_{\tau}$ & 0.19702; 0.19796 &  & 
\end{tabular}
\end{center}
\caption {\footnotesize Results analogous to those of Table \ref{tabl1},
but for the case $d_3^{(0)}\!=\!15$.
The results $r_{\tau}$ are given for the choices 
$\alpha_s(m^2_{\tau}; {\overline {\rm MS}})\!=\!0.331$ and $0.332$.
The central measured value $r_{\tau}\!=\!0.1960$ (\ref{rtauexp})
is achieved by the arithmetic average ${\bar r}_{\tau}$ at 
$\alpha_s(m^2_{\tau}; {\overline {\rm MS}})\!=\!0.3299$.
}
\label{tabl2}
\end{table}

\begin{table}[ht]
\par
\begin{center}
\begin{tabular}{c c  c  c}
($j$, $k$); approximant & 
$r_{\tau}$ [$\alpha_s(m^2_{\tau})\!=\!0.331; 0.332$] &
${\tilde c}$ &  comments \\
\hline \hline
$(1,3)$; $[2/1]$ & 0.19716; 0.19823 & +1.02 & $w^{\rm pole} \approx +1.$ \\
\hline
$(1,4)$; $[2/1]$ & 0.19780; 0.19888 & +1.37 & $w^{\rm pole} \approx +0.6$ \\
\hline\hline
$(1,3)$; TPS & 0.19744; 0.19852 & +0.025 & 
$\partial r_{\tau}/\partial {\tilde c} \approx -2.66 \cdot 10^{-3}$ \\
\hline
$(1,4)$; TPS & 0.19645; 0.19751 & +0.0 & 
$\partial r_{\tau}/\partial {\tilde c} \approx -2.92 \cdot 10^{-3}$ \\
\hline
$(2,3)$; TPS & 0.19603; 0.19708 & +0.375 & 
$\partial r_{\tau}/\partial {\tilde c} \approx -3.18 \cdot 10^{-3}$ \\
\hline
$(2,4)$; TPS & 0.19558; 0.19662 & +0.075 &
$\partial r_{\tau}/\partial {\tilde c} \approx -3.09 \cdot 10^{-3}$ \\
\hline
$(3,3)$; TPS & 0.19651; 0.19757 & -0.19 &
$\partial r_{\tau}/\partial {\tilde c} \approx -2.37 \cdot 10^{-3}$ \\
\hline
$(3,4)$; TPS & 0.19602; 0.19707 & -0.49 &
$\partial r_{\tau}/\partial {\tilde c} \approx -2.29 \cdot 10^{-3}$ \\
\hline\hline
arithm. average ${\bar r}_{\tau}$ & 0.19662; 0.19769 &  & 
\end{tabular}
\end{center}
\caption {\footnotesize Results analogous to those of Table \ref{tabl1},
but for the case $d_3^{(0)}\!=\!35$.
The results $r_{\tau}$ are given for the choices 
$\alpha_s(m^2_{\tau}; {\overline {\rm MS}})\!=\!0.320$ and $0.321$.
In the approach with the (${\rm N}^3 {\rm LO}$) TPS of
${\overline R}_j(w_k)$, the PMS condition (\ref{PMS})
is never exactly satisfied; there, the values of ${\tilde c}$
were chosen so that the (negative) slope 
$\partial r_{\tau}/\partial {\tilde c}$ is the least steep.
The central measured value $r_{\tau}\!=\!0.1960$ (\ref{rtauexp})
is achieved by the arithmetic average ${\bar r}_{\tau}$ at 
$\alpha_s(m^2_{\tau}; {\overline {\rm MS}})\!=\!0.3194$.
}
\label{tabl3}
\end{table}

\begin{table}[ht]
\par
\begin{center}
\begin{tabular}{c  c  c  c}
($j$, $k$); approximant & 
$r_{\tau}$ [$\alpha_s(m^2_{\tau})\!=\!0.329; 0.330$] &
${\tilde c}$ &  comments \\
\hline \hline
$(2,4)$; $[2/1]$ & 0.19521; 0.19613 & +0.01 & $w^{\rm pole} \approx -1.$ \\
\hline\hline
$(3,3)$; $[2/1]$ & 0.19656; 0.19750 & +0.707 & $w^{\rm pole} \approx +1.$ \\
\hline
$(3,3)$; $[2/1]$ & 0.19770; 0.19866 & +1.25 & $w^{\rm pole} \approx +1.$ \\
\hline\hline
$(2,3)$; TPS & 0.19668; 0.19763 & +1.1 & local max. \\
\hline
$(2,4)$; TPS & 0.19633; 0.19727 & +0.8 & local max. \\
\hline
$(3,3)$; TPS & 0.19672; 0.19767 & +1.05 & local max. \\
\hline
$(3,4)$; TPS & 0.19638; 0.19732 & +0.75 & local max. \\
\hline\hline
arithm. average ${\bar r}_{\tau}$ & 0.19651; 0.19745 &  & 
\end{tabular}
\end{center}
\caption {\footnotesize Results analogous to those of Table \ref{tabl1},
but for the different RSch: $c_2\!=\!c_1^2\!+\!3.4$, $c_k\!=\!c_1^k$
($k\!=\!3,4,\ldots$). 
Expression (\ref{rtaugen}) is used to calculate $r_{\tau}$.
RScl parameter is $\xi^2\!\equiv\!\mu^2/Q^2 = 1$
[$Q^2\!=\!m^2_{\tau} \exp( {\rm i} y)$]; $d_3^{(0)}\!=\!25$.
The results $r_{\tau}$ are given for the choices 
$\alpha_s(m^2_{\tau}; {\overline {\rm MS}})\!=\!0.329$ and $0.330$.
The central measured value $r_{\tau}\!=\!0.1960$ (\ref{rtauexp})
is achieved by the arithmetic average ${\bar r}_{\tau}$ at 
$\alpha_s(m^2_{\tau}; {\overline {\rm MS}})\!=\!0.3285$.
}
\label{tabl4}
\end{table}

\begin{table}[ht]
\par
\begin{center}
\begin{tabular}{c  c  c  c}
($j$, $k$); approximant & 
$r_{\tau}$ [$\alpha_s(m^2_{\tau})\!=\!0.327; 0.328$] &
${\tilde c}$ &  comments \\
\hline \hline
$(1,3)$; $[2/1]$ & 0.19478; 0.19570 & +1.50 & $w^{\rm pole} \approx -1.$ \\
\hline
$(1,4)$; $[2/1]$ & 0.19453; 0.19545 & +1.40 & $w^{\rm pole} \approx -1.$ \\
\hline
$(2,3)$; $[2/1]$ & 0.19582; 0.19676 & +1.11 & $w^{\rm pole} \approx -1.$ \\
\hline
$(2,4)$; $[2/1]$ & 0.19558; 0.19652 & +1.01 & $w^{\rm pole} \approx -1.$ \\
\hline
$(3,3)$; $[2/1]$ & 0.19654; 0.19749 & +0.80 & $w^{\rm pole} \approx -1.$ \\
\hline
$(3,4)$; $[2/1]$ & 0.19628; 0.19723 & +0.70 & $w^{\rm pole} \approx -1.$ \\
\hline\hline
$(1,3)$; $[2/1]$ & 0.19456; 0.19549 & +2.77 & $w^{\rm pole} \approx +1.$ \\
\hline
$(1,4)$; $[2/1]$ & 0.19538; 0.19632 & +3.07 & $w^{\rm pole} \approx +0.6$ \\
\hline
$(2,3)$; $[2/1]$ & 0.19612; 0.19708 & +2.25 & $w^{\rm pole} \approx +1.$ \\
\hline
$(2,4)$; $[2/1]$ & 0.19708; 0.19805 & +2.56 & $w^{\rm pole} \approx +0.6$ \\
\hline
$(3,3)$; $[2/1]$ & 0.19695; 0.19792 & +1.90 & $w^{\rm pole} \approx +1.$ \\
\hline
$(3,4)$; $[2/1]$ & 0.19796; 0.19894 & +2.205 & $w^{\rm pole} \approx +0.6$ \\
\hline\hline
$(1,3)$; TPS & 0.19400; 0.19492 & +1.9 & local max. \\
\hline
$(1,4)$; TPS & 0.19383; 0.19475 & +1.5 & local max. \\
\hline
$(2,3)$; TPS & 0.19528; 0.19622 & +1.45 & local max. \\
\hline
$(2,4)$; TPS & 0.19505; 0.19598 & +1.1 & local max. \\
\hline
$(3,3)$; TPS & 0.19602; 0.19696 & +1.1 & local max. \\
\hline
$(3,4)$; TPS & 0.19576; 0.19670 & +0.8 & local max. \\
\hline\hline
arithm. average ${\bar r}_{\tau}$ & 0.19564; 0.19658 &  & 
\end{tabular}
\end{center}
\caption {\footnotesize Results analogous to those of Table \ref{tabl4},
but for the RSch: $c_3\!=\!c_1^3\!+\!31.0$, $c_k\!=\!c_1^k$
($k\!=\!2,4,5,\ldots$). 
Expression (\ref{rtaugen}) is used to calculate $r_{\tau}$.
RScl parameter is $\xi^2 = 1$; $d_3^{(0)}\!=\!25$.
The results $r_{\tau}$ are given for the choices 
$\alpha_s(m^2_{\tau}; {\overline {\rm MS}})\!=\!0.327$ and $0.328$.
The central measured value $r_{\tau}\!=\!0.1960$ (\ref{rtauexp})
is achieved by the arithmetic average ${\bar r}_{\tau}$ at 
$\alpha_s(m^2_{\tau}; {\overline {\rm MS}})\!=\!0.3274$.
}
\label{tabl5}
\end{table}

\end{document}